# Robust estimation for small domains in business surveys

Paul A. Smith[1], Chiara Bocci[2], Nikos Tzavidis[1], Sabine Krieg[3], Marc J.E. Smeets[3]

[1] S3RI and Department of Social Statistics & Demography, University of Southampton, Highfield, Southampton, SO17 1BJ, UK.
[2] Department of Statistics, Computer Science, Applications "G. Parenti", University of Florence, Viale Morgagni 59, 50134 Firenze, Italy.
[3] Statistics Netherlands, Process Development & Methodology, P.O. Box 4481, 6401CZ Heerlen, The Netherlands.

**Abstract**

Small area (or small domain) estimation is still rarely applied in business statistics, because of challenges arising from the skewness and variability of variables such as turnover. We examine a range of small area estimation methods as the basis for estimating the activity of industries within the retail sector in the Netherlands. We use tax register data and a sampling procedure which replicates the sampling for the retail sector of Statistics Netherlands' Structural Business Survey as a basis for investigating the properties of small area estimators. In particular, we consider the use of the EBLUP under a random effects model and variations of the EBLUP derived under (a) a random effects model that includes a complex specification for the level 1 variance and (b) a random effects model that is fitted by using the survey weights. Although accounting for the survey weights in estimation is important, the impact of influential data points remains the main challenge in this case. The paper further explores the use of outlier robust estimators in business surveys, in particular a robust version of the EBLUP, M-regression based synthetic estimators, and M-quantile small area estimators. The latter family of small area estimators includes robust projective (without and with survey weights) and robust predictive versions. M-quantile methods have the lowest empirical mean squared error and are substantially better than direct estimators, though there is an open question about how to choose the tuning constant for bias adjustment in practice. The paper makes a further contribution by exploring a doubly robust approach comprising the use of survey weights in conjunction with outlier robust methods in small area estimation.

**Key words**

Small area estimation, M-quantile regression, bias adjustment, outliers

## 1. Introduction

Small area estimation encompasses a wide range of methods (Rao & Molina 2015) that have become a major component in the official statistician's toolkit, with many applications to a wide range of variables and domains in many different types of surveys. (We will use the terms domain and area interchangeably.) Nevertheless, the application of small area estimation methods to variables derived from business surveys has proved to be more challenging, because of the differences in the types of sample designs used in business surveys (Rivière 2002): skewed variables, detailed stratification, non-negligible sampling fractions, large variations in estimation weights, etc. Hidiroglou & Smith (2005) discuss some of the challenges in applying small area estimation in business surveys, and Burgard *et al*. (2014) demonstrate the effects of different sample designs as used in business surveys on the effectiveness of small area estimation methods. There are also few practical applications, and these are restricted to estimating proportions (Chandra *et al*. 2012) or numbers (Militino *et al*. 2015) of businesses with particular characteristics. Numbers and proportions work better with small area estimates in business surveys, as they are based on quantities whose underlying distributions are not

(or at least much less) skewed. It is therefore easier (though sometimes still challenging) to construct a suitable model in these cases than when dealing with numeric variables such as turnover or employment, which may vary by several orders of magnitude. More recently Ferrante & Pacei (2017) have extended Fay-Herriot (area-level) models to deal with skewed data and Fabrizi, Ferrante & Trivisano (2018) have used Bayesian procedures applied to area-level models for the same purpose.

In this paper, however, we consider the use of unit-level rather than area-level models as the basis of small area estimation, for estimation of turnover in small domains from the retail sector of the Structural Business Survey (SBS) in the Netherlands. The retail sector is one of those where small area estimation is likely to be most practical within business surveys, since there are generally very many retail businesses (and outlets), so that there is enough data to fit and compare models. Contrast this with industrial activities such as shipbuilding, which in many countries include only a handful of businesses; they therefore suffer from small numbers of observations from which to build models as well as all the difficulties caused by the structural characteristics of business surveys.

The skewness of distributions of business survey variables means that samples are likely to include outliers. These outliers in turn have a large effect on the estimation of population quantities which require special treatment (Chambers 1986), and they have an even larger effect in small area (domain) estimation (SAE), where sample sizes are considerably smaller and model-dependent estimation is the norm. This problem remains when the small area estimator is an indirect one, e.g. an empirical best linear unbiased predictor (EBLUP), since the weights, which determine the balance between use of information from within and beyond the domain, show that data from the small area of interest will still be used. Also, the estimates of the model parameters underpinning the estimator will themselves be affected by the sample outliers. In order to address these challenges, Chambers *et al*. (2014) considered the ways in which robust survey estimation procedures could be adapted to small area estimation.

The first approaches to explicitly address the issue of outlier robustness for small area estimation use plug-in robust prediction. That is, they replace the parameter estimates in optimal, but outlier sensitive, predictors by outlier robust versions (Chambers *et al*. (2014) call this a *robust projective* approach). Chambers and Tzavidis (2006) proposed an approach based on fitting outlier robust M-quantile models to the survey data, and Sinha & Rao (2009) proposed outlier robust procedures for small area estimation with linear random effects models. These robust projective approaches aim to produce a low prediction variance, but may involve an unacceptable prediction bias because they assume that non-sampled values in the target population are drawn from a distribution with the same properties as the sample non-outliers – in particular that prediction errors cancel on average in small areas. To avoid this bias Chambers *et al*. (2014) consider a *robust predictive* approach where a bias correction is applied through the use of two influence functions – the first quite restrictive to robustly choose a working model uninfluenced by the outliers, and a second, less restrictive, one to produce a bias correction.

Some of these methods have been evaluated in empirical work by Chambers *et al*. (2014). In this application to the retail sector in the Netherlands we use tax data to provide a known population which can be used to evaluate the methods, and demonstrate how they fare in a real data situation. This is consistent with a design-based evaluative framework as recommended by Tzavidis *et al*. (2018). Krieg *et al*. (2012) undertook a simulation study using tax data to approximate the population and auxiliary variables of the SBS, focussing on the situation where there are no strong auxiliary variables for individual units. They compared estimates from a direct estimator, the generalised regression estimator (GREG), with small area estimates from the EBLUP. The EBLUP was found to be biased because of the skewed distribution of turnover and the consequent prevalence and impact of outliers. Krieg *et al*. (2012) considered modelling transformations, albeit not data driven (see Rojas-Perilla *et*

*al*., 2019), of the turnover variable to deal with the skewness. This reduced the bias and the influence of the skewness on the resulting estimates, but increased their variability, such that the net effect was only a small improvement over the EBLUP.

In this paper we continue the investigation of the retail sector from the Dutch SBS, using tax data in different years as explanatory and outcome variables with samples based on the SBS sample design. The availability of information on the whole population of businesses from this administrative source allows us to evaluate the performance of different modelling methods against the true values. To compensate for the skewness of the turnover data, we consider models which are robust to unusual observations, comparing a variety of approaches which have been proposed in the literature. Of particular interest is the use of outlier robust methods in conjunction with survey weighting. This joint approach may provide double robustness, but possibly at the cost of increased variance. The structure of the remainder of the paper is as follows. In Section 2 we describe the pseudo-SBS data which form the basis of the comparisons, and in Section 3 the range of estimation methods to be compared is presented. Section 4 describes the results of applying these methods in the retail sector of the SBS. Section 5 makes an assessment of the sensitivity of the results to different aspects of the data and methods, and we draw conclusions and discuss the general applicability of our findings in Section 6.

## 2. Tax-based population and pseudo-Structural Business Survey data

The data to be modelled are derived from Value Added Tax (in Dutch BTW - belasting toegevoegde waarde) data of enterprises classified in the retail sector. These administrative data are provided to Statistics Netherlands to support the construction of the business register and survey-taking. The tax-turnover data are strongly correlated ($\rho \approx 0.7$, see Table 1) with the turnover which is the actual target of the Structural Business Survey, an annual survey undertaken by Statistics Netherlands. By using the tax-turnover we have information for almost the whole population (which we treat as if it is the whole population), and therefore can assess the outputs of the methods against the true population value. For this study, any enterprises which do not have tax data available for both periods considered are removed. The data in this study are taken from 2006 and 2007, with the first year's data playing the role of the business register, and the second year's data playing the role of a survey response, which is therefore available for the whole of the considered population. The correlation between the 2006 and 2007 tax-turnover is larger than between the turnover and tax-turnover (Table 1), but otherwise the data structure is very similar to the SBS.

**Table 1**: Correlation between tax in 2006 and 2007 and turnover

|          | Tax 2007 | turnover |
|----------|----------|----------|
| Tax 2006 | 0.996    | 0.732    |
| Tax 2007 |          | 0.727    |

The SBS has a stratified design, with strata defined by a combination of industrial classification according to NACE[1] (in twenty classes in the retail sector) and nine size classes. The largest businesses (in size classes 6-9, with employment 50 or greater) are completely enumerated. Since their information is in concept completely known, they are not included in the current study. The sample sizes in other strata were determined based on Neyman allocation with some additional constraints on subpopulations (including for the retailing sector), and this reflects the design of the 2009 SBS. Within strata, samples are selected from the population of enterprises by simple random sampling without replacement. We have population size $N$ = 63958, five size classes and 20 industries, and will denote the size class × industry strata by $h$; the total sample size is 5074, with sample sizes per industry varying from 21 to 769. The design of the SBS used in this study is described in Krieg *et al*. (2012, Appendix A) – the smallest enterprises have the smallest inclusion probabilities and the largest weight, and there are also quite large differences in weights between industries.

---
[1] Nomenclature statistique des activités économiques dans la Communauté européenne, the standard European industrial classification.

Repeated samples were drawn from the population of tax-turnover data using the design of the SBS in the retail sector, and inclusion probabilities were retained with the data. Since the population is fixed, the simulation is over repeated sampling, and not over repeated realisations of a population from a model, and we are therefore assessing the design-based properties of these small area estimators as recommended by Tzavidis *et al*. (2018).

Four auxiliary variables are available with which to construct models to predict the 2007 tax-turnover:
- tax-turnover from 2006, *tax1*
- industrial classification, *ind*
- size class, based on the working persons in the business in bands 1, 2-4, 5-9, 10-19, 20-49, *sc*
- employment, that is the number of working persons, *wp*

and we additionally denote the tax-turnover to be estimated (for 2007) as *tto*, and the design weight as $d_i$ for business *i*. It is slightly curious to have two predictors derived from the same working persons information – both the size class and employment in its own right – and Statistics Netherlands' preferred models (Krieg *et al*. 2012) include *both* variables. We will consider whether this approach, which implies an element of nonlinearity in the effect of working persons, is a valuable strategy during our model fitting.

For the purposes of evaluation we will consider the small areas to be the 20 industry classes within the retail sector, *ind* = 1, …, 20. In line with common practice in business surveys we will estimate totals for *tto* within the domains; (the equations for estimators in this paper are therefore for totals, in contrast to the more usual formulation for estimation of means). There are two potential challenges in this situation. First, the sampling weights of observations within the industry classes vary, because selected businesses come from different original strata *h*. The sampling weights are usually assumed to be ignorable given the model in model-based estimation (which includes small area estimation), but the large variation in the sampling weights in business surveys makes this assumption problematic. The investigations below are assessing the effect of ignoring the weights as well as the effect of outliers among the responses, and we further consider whether using the weights is helpful in section 5.4. The second challenge is that when the survey is undertaken, some businesses may be found to have changed their size and/or industry classification; we do not consider this situation in this investigation. If we did wish to make estimates for domains defined by the new information, we would have less information on the domain sizes in the population, and therefore more variable estimators.

## 3. The small area methods

### 3.1 Direct estimation

Several different estimators are compared in this study. A natural comparator to use as a baseline for evaluation is the direct estimator which is often preferred by National Statistical Institutes (NSIs) because of its unbiasedness and objectivity (Rao 2011). The simplest of these is the Horvitz-Thompson (HT) estimator:

$$\hat{y}_{ind}^{HT} = \sum_{i \in s \cap ind} d_i y_i \tag{1}$$

where the $d_i = 1/\pi_i$ are the inverse of the selection probabilities $\pi_i$. Auxiliary information is typically very valuable in estimation in business surveys, and although the HT estimator is used, it is unusual; therefore we also consider the generalised regression (GREG) estimator (Särndal *et al*. 1992, chapter 6):

$$\hat{y}_{ind}^{GREG} = \hat{y}_{ind}^{HT} + \hat{\boldsymbol{\beta}}^{GREG}\left[\mathbf{X}_{ind} - \hat{\mathbf{x}}_{ind}^{HT}\right] \qquad (2)$$

where $\mathbf{X}_{ind}$ is a vector of known totals of auxiliary variables and $\hat{\mathbf{x}}_{ind}^{HT}$ is the vector of HT estimates of these variables from the sample calculated using (1) with $y_i$ replaced by the appropriate auxiliary variables. There are many possible models, derived from suitable choices of the definition of **X** and **β**, but in this case we use the actual approach from the Dutch SBS with auxiliary variables defined by *tax*1 × *ind* × size class group, where the size class group is an aggregation of *sc* into 1-9 and 10-49 working persons. We note that it is possible to write such estimators in the same form as (1) as $\hat{y}_{ind}^{GREG} = \sum_{i \in s \cap ind} w_i y_i$ with modified weights $w_i$, and that domain estimation can then be used to obtain the required industry level estimates. Of course, this approach fails in the case that there are no sample units in the domain of interest, but the sample design ensures that this does not happen in our example.

3.2  Non-robust small area methods

The basis of small area estimation is to borrow strength across a wider pool of data than the domain in question (*indirect* estimators). In general this produces biased estimators, in exchange for a substantial reduction in the variance, with the trade-off between bias and variance chosen through the mean squared error. The classical approach when the microdata are available for modelling is to assume that the population data follow a random effects model (Battese, Harter & Fuller 1988, Rao & Molina 2015 section 4.3), and in this case the structure of the data with enterprises within industries suggests a two-level random effects model:

$$\mathbf{y} = \mathbf{X}\boldsymbol{\beta} + \mathbf{Z}\mathbf{u} + \mathbf{e} \qquad (3)$$

where **β** contains the fixed effects, the rows of **X** contain the corresponding auxiliary data, the rows of **Z** are dummy variables for industry, $\mathbf{u} \sim N(\mathbf{0}, \boldsymbol{\Sigma}_u)$ is a vector of industry-specific random effects and $\mathbf{e} \sim N(\mathbf{0}, \boldsymbol{\Sigma}_e)$ is a vector of individual-specific random effects. We have followed Krieg *et al*. (2012) who in their study of the effect of transformations of the outcome variable considered a model using all the available variables

$$tto_{i,ind} = \beta_0 + \beta_1 tax1_{i,ind} + \boldsymbol{\beta}_2 \mathbf{sc}_{i,ind} + \beta_3 wp_{i,ind} + \beta_4 \left(tax1 \times wp\right)_{i,ind} + u_{ind} + e_{i,ind} . \qquad (4)$$

and this is the basis of the main results. Note that this model contains both the working persons and the size classes derived from the same variable. We also explore the effect of using a smaller model using only the size class variable without the more detailed information on number of working persons:

$$tto_{i,ind} = \beta_0 + \beta_1 tax1_{i,ind} + \boldsymbol{\beta}_2 \mathbf{sc}_{i,ind} + u_{ind} + e_{i,ind} \qquad (5)$$

This produces only small changes in the results (section 5.3), but the AIC for (4) is lower (see table S1 in the supplementary material for model fit statistics), which suggests that there is a nonlinear effect of size class so that the incorporation of both *wp* and the *sc* variable derived from it is beneficial. We present detailed results from model (5) in the on-line supplementary material, but the results are not qualitatively different from model (4).

Note that (3) does not incorporate the design weights $d_i$, in contrast to (1) and (2). Including the variables (*ind*, *sc*) which explain the differences in the weights as predictors is one strategy to reduce the impact of the informative design (Pfeffermann 2011). We will assume that the sampling is approximately ignorable and the weights are not needed if these variables are included in the model, though inasmuch as this is not true it will have an impact on the repeated sampling results in section 4. However, one of the main features that distinguish business surveys is the importance of the weights to deal with differential sampling due to size differences. We therefore consider weighted versions of several methods to assess whether including the weights improves the properties of the small area estimates (see for example the end of this subsection).

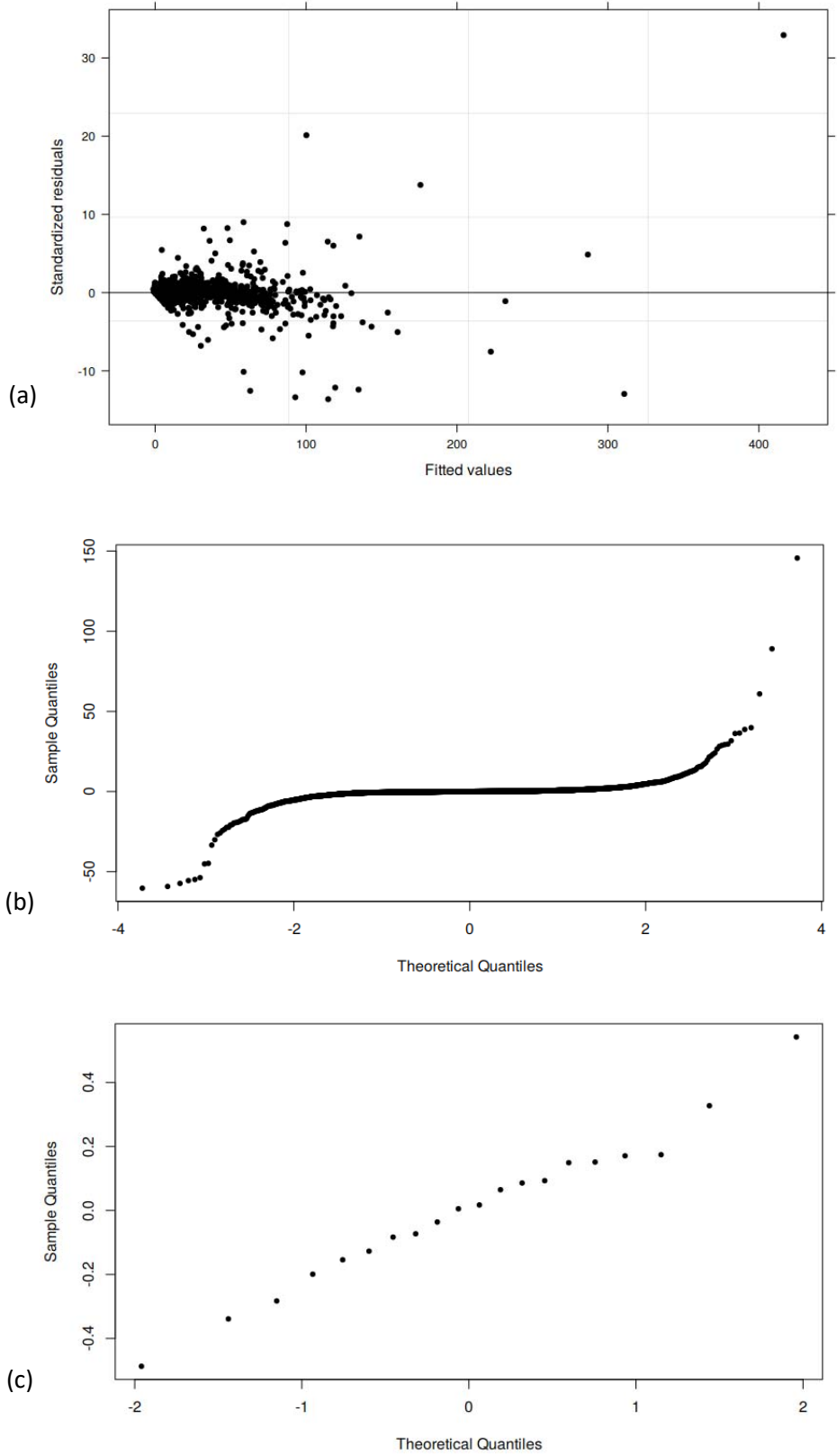

Fig. 1: Diagnostic plots from model (3) fitted to one sample from the SBS retail population. (a) Standardised residuals vs fitted values; (b) Normal probability plot for enterprise-level residuals; (c) Normal probability plot for industry-level residuals.

Let $\hat{\boldsymbol{\beta}}$ and $\hat{\mathbf{u}}$ denote estimates of the fixed and random effects in (3). The EBLUP of the industry *ind* total of the $y_i$ (= *tto*$_j$ in our case) under (3) is then

$$\hat{y}_{ind}^{EBLUP} = \sum_{i \in ind \cap s} y_i + \sum_{i \in ind \cap r} \left( \mathbf{x}_i^T \hat{\boldsymbol{\beta}} + \mathbf{z}_i^T \hat{\mathbf{u}} \right) \qquad (6)$$

where $\hat{\mathbf{u}}$ denotes the vector of the estimated area-specific random effects and we use *s* and *r* to denote sample and non-sample units respectively.

It is important to check the diagnostics from fitting the assumed population model (3) to the sample data, specifically residual and normal probability plots, to ensure that the model assumptions are satisfied, before the model can safely be used for prediction. Diagnostics from one realised sample are shown in Fig. 1.

There are clearly some large and influential outliers in the residuals in Fig 1a. The industry-level errors are plausibly normal in Fig. 1c, particularly considering the small number of observations. However, it is clear that the distribution of the enterprise-level residuals in Fig. 1b departs substantially from normality, and this is a characteristic problem when applying such methods in business surveys, deriving from the skewed distributions of the data (Rivière 2002).

Krieg *et al.* (2012) explored models fitted to transformations of *tto*; in business surveys the variance often increases with the size of the business (which also looks true from inspection of Fig. 1a), so log or root transformations are often appropriate to stabilise variances. Krieg *et al.* discover that using the square root transformation of *tto* in (3) results in negatively skewed residuals, and find that *tto*$^{1/3}$ produces approximately normal residuals. Nevertheless, following the approach for small area estimation for transformed variables developed by Chandra & Chambers (2011) based on *tto*$^{1/3}$ does not substantially reduce the mean squared error, which seems still to be affected by outliers after the transformation.

So one option is to extend the method to allow the enterprise-level variance to vary as a function of the size class by replacing $\mathbf{e}$ in (3) by $\mathbf{e}_{sc} \sim N\left(\mathbf{0}, \boldsymbol{\Sigma}_{e,sc}\right)$, or as a function of the number of working persons. In fact using *wp*$^2$ provides a better characterisation of the variance than *wp* (see table S1 in the supplementary material for model fit statistics), so in this case replace $\mathbf{e}$ in (3) by $\mathbf{e} = \left(\mathbf{wp}^2\right)^T \boldsymbol{\varepsilon}$ with $\boldsymbol{\varepsilon} \sim N\left(\mathbf{0}, \boldsymbol{\Sigma}_\varepsilon\right)$. These reduce the impact of outliers in larger enterprises by allowing for larger variances in larger size classes, but are not fully robust to outliers. In the results below, allowing the variance to change as a function of the size class gives smaller relative root mean square errors, so we restrict our presentation to this version.

Another way to improve the fit of the model is to account for the weighting in fitting the random effects model, which is a quite often used in modelling business survey data. This approach also bridges the two schools of thought in survey estimation, namely design-based and model-based estimation. One strategy to account for the weighting is to use the approach of You & Rao (2002), hereafter referred to as the pseudo-EBLUP,

$$\hat{y}_{ind}^{pEBLUP} = \hat{\gamma}_{ind,w} \sum_{i \in s_{ind}} w_{ind,i} y_{ind.i} + \left( \mathbf{X}_{ind} - \hat{\gamma}_{ind,w} \sum_{i \in s_{ind}} w_{ind,i} \mathbf{x}_{ind.i} \right) \hat{\boldsymbol{\beta}}_w \qquad (7)$$

where $\hat{\gamma}_{ind,w}$ is the shrinkage factor which defines the weights for the direct and regression synthetic parts of the estimator (for full details see You & Rao 2002).

### 3.3 Robust projective methods

Our analysis so far indicates that the impact of influential data points on the model fit is significant and thus it cannot be alleviated by just extending the structure of the random effects model. In our view the use of outlier robust fitting methods is essential when working with business survey data. In the last decade there has been a series of developments on outlier robust small area estimation, which we will summarise in the next sections.

The natural approach to outlier robust small area estimation is to extend the random effects model fit to control for the impact of outliers. Using work by Richardson & Welsh (1995) and Welsh & Richardson (1997), Sinha & Rao (2009) developed a robust EBLUP (REBLUP) estimator, by replacing the usual maximum likelihood equations with alternative versions which reduce the effect of outliers using a Huber function

$$\psi(a) = a \min(1, b_\psi / |a|) \qquad (8)$$

(Huber 1973) where $b_\psi > 0$ is a tuning constant. For residuals with a normal distribution, the theoretical value $b_\psi = 1.345$ is generally considered to be optimal (derived from Holland & Welsch 1977), and we follow this strategy here, as it is the standard approach. Nevertheless, for skewed distributions such as those found in business surveys data other values may be more suitable (Dawber *et al.* in prep.). This function is applied to the model residuals, scaled by the inverse of their variances and leads to robust estimates of the parameters in (3) $\hat{\boldsymbol{\beta}}^\psi$ and $\hat{\mathbf{u}}^\psi$ where the $\psi$ superscript denotes the dependence of the fitted parameters on the chosen influence function. Using these parameters in the EBLUP derivation gives us the REBLUP estimates of the industry totals

$$\hat{y}_{ind}^{REBLUP} = \sum_{i \in ind \cap s} y_i + \sum_{i \in ind \cap r} \left( \mathbf{x}_i^T \hat{\boldsymbol{\beta}}^\psi + \mathbf{z}_i^T \hat{\mathbf{u}}^\psi \right) \qquad (9)$$

(for details of the derivation see Chambers *et al.* 2014).

Applying the REBLUP estimator with $b_\psi = 1.345$ to the SBS retail data has some interesting effects – the robustly estimated industry-level effect resulted in convergence problems because the between-industry variance component is close to the boundary of the parameter space. So the industry-level effect apparent in the EBLUP (and EBLUP with transformed data) is due entirely to the effect of the outliers. If we remove this term from the regression, we return to a standard linear model for which the REBLUP methodology is not appropriate and which leads back to a synthetic estimator. In this approach our assumed population model (3) and small area estimator (6) is clearly not appropriate based on the sample data. So if we want to use (3) as a basis for small area estimation in this example we must examine alternative robust estimation methods.

A second, simple robust projective approach is to use M-regression, which uses a Huber function (8) to reduce the influence of outlying residuals in model fitting. The resulting robust estimate of the regression parameters $\hat{\boldsymbol{\beta}}^{rob,\psi}$, derived by an iterative fitting process (Draper & Smith 2014, pp569-572) can be used to calculate a simple robust synthetic estimator

$$\hat{y}_h^{RSYN} = \sum_{i \in h \cap s} y_i + \sum_{i \in h \cap r} \mathbf{x}_i^T \hat{\boldsymbol{\beta}}^{rob,\psi} \qquad (10)$$

Note that M-regression contains only fixed effects; we use the same fixed effects as in model (4) in the paper (and (5) in the supplementary material); an industry fixed effect is not added, so for this model the weights are expected not to be ignorable. However, we also expect that as a result of reducing the impact of outliers, the importance of weighting will be reduced.

A further robust projective approach is the M-quantile regression-based method described by Chambers and Tzavidis (2006). M-quantile regression is a robust regression approach based on an influence function similar to (8), which allows us to fit a model which produces an optimal mixture of regressions for the mean and median, preserving some of the sensitivity of the mean while taking advantage of the robustness of the median (Breckling & Chambers 1988). We again use $b_\psi = 1.345$.

Chambers & Tzavidis's approach is based on a linear model for the M-quantile regression of **y** on **X**,
$$m_q(\mathbf{X}) = \mathbf{X}\boldsymbol{\beta}_q^\psi$$
where $m_q(\mathbf{X})$ denotes the M-quantile of order $q$ of the conditional distribution of $y$ given **X**. An estimate $\hat{\boldsymbol{\beta}}_q^\psi$ of $\boldsymbol{\beta}_q^\psi$ can be calculated for any value of $q$ in the interval (0,1), and for each unit in sample we define its unique M-quantile coefficient under this fitted model as the value $q_i$ such that $y_i = \mathbf{x}_i^T \hat{\boldsymbol{\beta}}_{q_i}^\psi$. Then for each small domain *ind*, we calculate the sample average of these coefficients and denote it by $\bar{q}_{ind} = \frac{1}{n_{ind}} \sum_{i \in ind} q_i$. The M-quantile estimate of the total of $y$ in industry *ind* is then

$$\hat{y}_{ind}^{MQ} = \sum_{i \in ind \cap s} y_i + \sum_{i \in ind \cap r} \mathbf{x}_i^T \hat{\boldsymbol{\beta}}_{\bar{q}_{ind}}^\psi . \tag{11}$$

Note that the M-quantile approach to small area estimation allows for different predicted values in different industries. This is in contrast to the robust synthetic estimator (10). We believe this approach to quantifying between industry variability may overcome the numerical issues we encountered when fitting the outlier robust version of the random effects model that resulted in a zero between-industry variance component. The robust projective approaches are called naïve estimators by Tzavidis *et al*. (2010), and we will refer to (11) as the naïve M-quantile estimator. This is because the working model, which is equivalent to an outlier rejection approach, is projected onto the non-sampled part of the population without accounting for the possibility that outliers also exist in this part too. Although this approach will reduce the variance of the estimates, we expect that it will also introduce bias in the small area estimates. The assumption that all the outliers are observed in the sample is a strong one. Methods that relax this assumption are presented in the next section.

As in the case of the EBLUP, we may be interested in using a weighted version of the M-quantile estimator. Arguing for design-consistency, Fabrizi *et al*. (2014) proposed weighted versions of M-quantile estimators, which we also include in our empirical assessments as we are interested in studying the use of survey weights in conjunction with outlier robust estimation, see section 5.4. The weighted version of the naïve M-quantile estimator is

$$\hat{y}_{ind}^{wMQ} = \sum_{i \in ind \cap s} y_i + \sum_{i \in ind \cap r} \mathbf{x}_i^T \boldsymbol{\beta}_{w,\bar{q}_{ind}}^\psi \tag{12}$$

where the $w_i$ are the sampling weights and $\boldsymbol{\beta}_{w,\bar{q}_{ind}}^\psi$ is also estimated using the sampling weights, see equation (14) in Fabrizi *et al*. (2014).

### 3.4 Robust predictive methods

The naïve robust projective approaches of section 3.3 assume that all the non-sampled units follow the (robustly fitted) working model, whereas in practice there will in general be businesses like the outliers among the non-sampled units. Hence, using the M-quantile predictions for the out-of-sample units directly as in (11) leads to a biased estimator of the population total in each small domain. Using the ideas in Chambers (1986), Tzavidis *et al*. (2010) substitute a consistent estimator of the distribution function, using the approach of Chambers & Dunstan (1986), to derive a version of the M-quantile estimator adjusted for bias, and we label this MQCD:

$$\hat{y}_{ind}^{MQCD} = \sum_{i\in ind\cap s} y_i + \sum_{i\in ind\cap r} \mathbf{x}_i^T \hat{\boldsymbol{\beta}}_{\bar{q}_{ind}}^{\psi} + \frac{N_{ind}-n_{ind}}{n_{ind}} \sum_{i\in ind\cap s} \left(y_i - \mathbf{x}_i^T \hat{\boldsymbol{\beta}}_{\bar{q}_{ind}}^{\psi}\right) \tag{13}$$

In principle, this estimator works by adding a third term to correct for the potential bias at the cost of allowing the variance to increase. It is easy to note that this estimator is a model-based version of a GREG estimator. Fabrizi *et al*. (2014) also propose a weighted version of (13):

$$\hat{y}_{ind}^{wMQCD} = \sum_{i\in ind\cap s} w_i y_i + \left(\sum_{i\in ind\cap U} x_i^T - \sum_{i\in ind\cap s} w_i x_i^T\right) \hat{\boldsymbol{\beta}}_{w,\bar{q}_{ind}}^{\psi} \tag{14}$$

with $\hat{\boldsymbol{\beta}}_{w,\bar{q}_{ind}}^{\psi}$ again estimated using equation (14) from their paper. The disadvantage of (13) and (14) is that we cannot control the impact of the third term, which can lead to a large increase in the variance.

To obtain an estimator which accounts for this possibility, we must reduce the effect of large residuals in the third term of (13), but by less than in the second term, so that we allow a part of their effect into the bias adjustment. A further Huber function $\phi$ is therefore needed (defined as in (8)), with a different tuning constant from $\psi$ such that $|\phi(\cdot)| \geq |\psi(\cdot)|$ (Tzavidis *et al*. 2010 p172), that is the tuning constant for $\phi$, $b_\phi$, takes larger values than $b_\psi$, and the estimation is less robust than for $\psi$. The (robust predictive) robustly bias-adjusted version of the M-quantile estimator (labelled MQWR because of its derivation from Welsh & Ronchetti (1998)) is:

$$\hat{y}_{ind}^{MQWR} = \sum_{i\in ind\cap s} y_i + \sum_{i\in ind\cap r} \mathbf{x}_i^T \hat{\boldsymbol{\beta}}_{\bar{q}_{ind}}^{\psi} + \frac{N_{ind}-n_{ind}}{n_{ind}} \sum_{i\in ind\cap s} \omega_{ind}^{MQ} \phi\left\{\frac{\left(y_i - \mathbf{x}_i^T \hat{\boldsymbol{\beta}}_{\bar{q}_{ind}}^{\psi}\right)}{\omega_{ind}^{MQ}}\right\}, \tag{15}$$

where $\omega_{ind}^{MQ}$ is a robust estimator of the scale of the residuals $y_i - \mathbf{x}_i^T \hat{\boldsymbol{\beta}}_{\bar{q}_{ind}}^{\psi}$ in domain *ind*. The tuning constant for the Huber function in the third term of (15) codifies the trade-off between bias and variance. We use several values of this tuning constant ($b_\phi$ = 1, 2, 3) to investigate its effect on the small domain estimates. Note that we are also testing values of $b_\phi < b_\psi$ in contradiction of the theory set out by Tzavidis *et al*. (2010). See also section 5.1 below. A similar robust predictive extension to the REBLUP estimator (9) is also available (Dongmo Jiongo *et al*. 2013), but since the REBLUP is not appropriate for our data we do not consider it.

### 4. Repeated sampling results

We start with the known retail sector business population derived from the tax data, and consider only size classes 1-5 since the others are completely enumerated and therefore fully known. We select 500 replicate samples using the sample design of the SBS. In some industries there remain some size classes which are completely enumerated, so some businesses are always included. In each of these samples we apply the set of estimators described in table 2, using model (4); we explore the sensitivity to the model choice in section 5.3 and the supplementary material.

The residuals from the fixed part of model (4) are shown in Fig 2(a). This shows that there are outliers, but the three labelled ones dominate measures of Cook's distance (Fig. 2(b)). Observation 50010 is in one of the completely enumerated strata, so it is included in all the replicate samples. See section 5.2 for more on the effect of these observations.

For each sample we calculate estimates for the small domains formed by the 20 industries, with population sizes ranging from 110 to 8407 and sample sizes from 21 to 769. We then calculate the per cent relative bias in each estimator $\left( \frac{100}{500 y_{ind}} \sum_{k=1}^{500} \hat{y}_{ind,k} \right) - 100$, where $\hat{y}_{ind}$ is any of the considered estimators and $y_{ind}$ the true value for industry *ind* calculated from the population, with *k* indexing the repeated samples. The per cent empirical relative root mean squared error (rrmse) is similarly calculated as $\frac{100}{y_{ind}} \left[ \frac{1}{500} \sum_{k=1}^{500} \left( \hat{y}_{ind,k} - y_{ind} \right)^2 \right]^{\frac{1}{2}}$. The results of the simulation for each industry *ind* are presented in Tables 3 and 4; Table S2 restates summary information for the full range of methods.

The interpretation of the results from Tables 3 and 4 (and S2 in the supplementary material) largely follows the expectations from the development of the theory in section 3, though with some interesting nuances. The HT estimator is approximately unbiased as anticipated, but in most industries with a very large relative rmse because of its high variance, which makes the estimates unusable. Introducing strong explanatory information through the GREG substantially improves the rrmse, and there is only a small increase in the relative bias resulting from the small sample sizes within industries. The EBLUP also reduces the rrmse, though not to an acceptable level, and is biased; in fact the GREG has better properties than the EBLUP in this example even though both are affected by outliers. Accounting for the variance structure in a more reasonable way in the EBLUP by allowing the variance to be different within different size classes (var=*f(sc)*) reduces the rrmse to less than half that of the HT estimator and improves somewhat on the GREG, though the estimator can be seen to have a small bias over repeated sampling.

Fig. 2: (a) Plot of residuals against fitted values for the whole population from the fixed part of model (4). The three points with (by some way) the largest values of Cook's distance are labelled. (b) Cook's distances for each observation (compare with Fig. S1(b)).

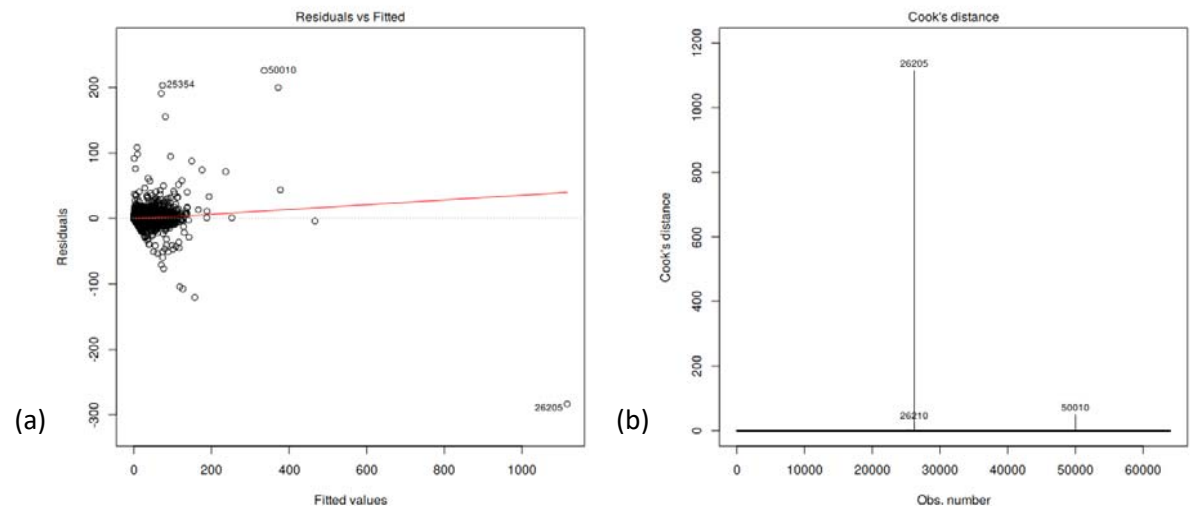

Moving to the robust synthetic estimator using parameters fitted with M-regression has a substantial further impact on the rmse, partly from a reduction in the bias and partly from reduced variance. In particular it reduces the rrmse in industries with large rrmse values under other methods considered so far. This suggests that using methods which are robust to the outliers in the data may have a substantial impact on the quality of the estimates. Most of the remaining larger rrmses are dominated by the bias, however, so further exploration of bias correction seems potentially valuable.

**Table 2:** Sequence of estimators considered for small area estimation; the estimators appear in the same order in Tables 3 and 4.

| Estimator | Symbol | Equation | Method information | | | |
|---|---|---|---|---|---|---|
| | | | Fixed effects only or random effects model? | Regression type | Weighted? | Outlier robust? |
| Direct (HT) | $\hat{y}_{ind}^{HT}$ | (1) | constant fixed | ordinary | no | no |
| Direct (GREG) | $\hat{y}_{ind}^{GREG}$ | (2) | fixed | ordinary | no | no |
| EBLUP | $\hat{y}_{ind}^{EBLUP}$ | (6) with variables from (3) with $\mathbf{e} \sim N(\mathbf{0}, \mathbf{\Sigma}_e)$ | random | ordinary | no | no |
| EBLUP (var = f(SC)) | $\hat{y}_{ind}^{EBLUP}$ | (6) with variables from (3) with $\mathbf{e} \sim N(\mathbf{0}, \mathbf{\Sigma}_{e,sc})$ | random | ordinary | no | no |
| pseudo-EBLUP | $\hat{y}_{ind}^{pEBLUP}$ | (7) | random | ordinary | yes | no |
| Robust synthetic | $\hat{y}_{h}^{RSYN}$ | (10) | fixed | M-regression | no | projective |
| M-quantile naïve ($b_\psi = 1.345$) | $\hat{y}_{ind}^{MQ}$ | (11) | fixed | M-quantile | no | projective |
| M-quantile bias-adjusted ($b_\psi = 1.345$) | $\hat{y}_{ind}^{MQCD}$ | (13) | fixed | M-quantile | no | predictive |
| M-quantile robustly bias-adjusted ($b_\psi = 1.345$, $b_\phi = 1, 2, 3$) | $\hat{y}_{ind}^{MQWR}$ | (15) | fixed | M-quantile | no | predictive |
| weighted M-quantile naïve ($b_\psi = 1.345$) | $\hat{y}_{ind}^{wMQ}$ | (12) | fixed | M-quantile | yes | projective |
| weighted M-quantile bias-adjusted ($b_\psi = 1.345$) | $\hat{y}_{ind}^{wMQCD}$ | (14) | fixed | M-quantile | yes | predictive |

**Table 3:** Industry-specific relative bias (rb) (%) of small area point estimators of total *tto*. The best-performing methods are indicated in bold in the summary lines.

| Industry | Direct | | EBLUP | EBLUP | pseudo-EBLUP | robust synthetic | M-quantile naive | M-quantile bias-adjusted | M-quantile robustly bias-adjusted | | | M-quantile naive | M-quantile bias-adjusted |
|---|---|---|---|---|---|---|---|---|---|---|---|---|---|
| | HT | GREG | | | | | | | unweighted | | | weighted | |
| | | | | var = f(sc) | | | | | $b_\phi = 1$ | $b_\phi = 2$ | $b_\phi = 3$ | | |
| 52110 | -0.28 | -0.03 | -1.03 | 0.26 | 0.02 | 1.01 | 0.64 | 0.06 | 0.39 | 0.38 | 0.42 | 0.32 | -0.05 |
| 52120 | -0.59 | -1.00 | 0.07 | 2.86 | 1.47 | 1.16 | 0.35 | 0.07 | 0.11 | -0.12 | -0.31 | 0.20 | -0.22 |
| 52200 | 0.08 | -0.04 | 0.79 | 1.91 | 0.37 | 1.24 | 0.29 | 1.42 | 0.00 | -0.12 | -0.15 | 0.07 | 0.04 |
| 52310 | -0.16 | -0.04 | -1.14 | -0.66 | -0.77 | -0.04 | 0.44 | 0.02 | 0.61 | 0.60 | 0.54 | 0.29 | -0.02 |
| 52321 | -0.27 | 0.11 | -0.62 | 2.50 | 1.85 | 2.71 | 1.70 | 0.23 | 1.26 | 1.24 | 1.26 | 1.48 | 0.21 |
| 52330 | -0.24 | -0.83 | -5.28 | 1.86 | 5.84 | 1.39 | 1.10 | -29.51 | 1.21 | 0.95 | 0.47 | 0.74 | 0.11 |
| 52413 | 0.48 | 0.26 | -5.16 | 5.10 | 8.44 | 5.38 | 2.15 | -10.33 | -0.10 | -1.91 | -3.38 | 1.97 | 0.21 |
| 52422 | -0.48 | -0.32 | 2.01 | -0.50 | -0.62 | -0.73 | -0.75 | 7.26 | -0.41 | -0.01 | 0.25 | -0.99 | -0.22 |
| 52431 | 0.19 | -0.04 | -2.35 | -0.92 | -1.07 | -1.44 | -0.81 | -0.22 | -0.52 | -0.22 | -0.03 | -1.04 | -0.50 |
| 52440 | 0.26 | 0.03 | -8.25 | 0.56 | 0.24 | 0.78 | 0.22 | -6.09 | -0.49 | -1.12 | -1.73 | -0.09 | 0.00 |
| 52450 | 0.56 | 3.00 | -1.46 | 1.59 | 1.60 | 2.41 | 1.70 | -9.98 | 1.55 | 1.54 | 1.53 | 1.28 | 0.12 |
| 52460 | 0.29 | 0.07 | -4.07 | -1.12 | -0.76 | -1.64 | -1.68 | -1.16 | -1.66 | -1.69 | -1.72 | -1.92 | 0.10 |
| 52470 | 0.06 | -0.26 | -2.89 | 1.75 | 0.94 | 2.42 | 1.52 | -2.24 | 1.30 | 1.23 | 1.18 | 1.22 | -0.14 |
| 52485 | 0.33 | 0.08 | -4.41 | -1.44 | -0.96 | -2.01 | -1.52 | 0.13 | -0.83 | -0.51 | -0.53 | -1.78 | 0.13 |
| 52491 | -0.01 | 0.31 | 2.24 | 0.54 | 0.33 | -0.39 | -0.11 | 4.71 | 0.33 | 0.73 | 1.05 | -0.37 | 0.04 |
| 52500 | 0.90 | 0.27 | 8.45 | 1.88 | 3.16 | -1.25 | -1.78 | 6.10 | -1.60 | -1.38 | -1.15 | -2.22 | 0.65 |
| 52610 | -0.68 | 1.16 | 21.38 | -3.57 | -2.87 | -5.08 | -4.58 | 31.29 | -4.20 | -3.88 | -3.61 | -4.92 | -0.06 |
| 52620 | 0.31 | 1.36 | -1.86 | 2.24 | 1.28 | 0.98 | 0.53 | 0.43 | 0.58 | 0.58 | 0.56 | 0.27 | 0.14 |
| 52630 | -0.48 | 0.01 | 0.26 | -0.46 | 0.08 | -0.52 | -1.27 | -0.82 | -1.29 | -1.25 | -1.21 | -1.66 | -0.06 |
| 52700 | 0.12 | -0.06 | -2.42 | -0.20 | 0.99 | -2.42 | -1.87 | 2.05 | -1.31 | -1.03 | -0.92 | -2.15 | -0.09 |
| median (rb) | 0.07 | 0.22 | -1.30 | 0.55 | 0.35 | 0.37 | 0.25 | 0.07 | -0.05 | -0.12 | -0.09 | **-0.01** | **0.02** |
| mean (rb) | **0.02** | 0.49 | -0.29 | 0.71 | 0.98 | 0.20 | -0.19 | -0.33 | -0.25 | -0.30 | -0.37 | -0.47 | **0.02** |
| mean abs(rb) | 0.34 | 0.93 | 3.81 | 1.60 | 1.68 | 1.75 | 1.25 | 5.71 | 0.99 | 1.02 | 1.10 | 1.25 | **0.16** |

**Table 4:** Industry-specific relative rmse (%) of small area point estimators of the total *tto*. The best-performing methods are indicated in bold in the summary lines.

| Industry | Direct | | EBLUP | EBLUP | pseudo-EBLUP | robust synthetic | M-quantile naïve | M-quantile bias-adjusted | M-quantile robustly bias-adjusted | | | M-quantile naïve | M-quantile bias-adjusted |
|---|---|---|---|---|---|---|---|---|---|---|---|---|---|
| | HT | GREG | | | | | | | unweighted | | | weighted | |
| | | | | var = $f(sc)$ | | | | | $b_\phi = 1$ | $b_\phi = 2$ | $b_\phi = 3$ | | |
| 52110 | 3.88 | 1.99 | 2.71 | 4.06 | 1.88 | 1.12 | 0.89 | 2.62 | 0.80 | 0.86 | 0.96 | 0.78 | 2.07 |
| 52120 | 10.22 | 3.31 | 2.51 | 4.60 | 2.76 | 1.67 | 1.60 | 3.29 | 1.62 | 1.68 | 1.75 | 1.60 | 3.24 |
| 52200 | 4.33 | 1.91 | 2.32 | 2.96 | 1.83 | 1.29 | 0.50 | 2.48 | 0.44 | 0.50 | 0.55 | 0.42 | 1.86 |
| 52310 | 2.98 | 1.01 | 1.99 | 2.04 | 1.96 | 0.39 | 0.62 | 0.75 | 0.74 | 0.75 | 0.72 | 0.53 | 1.00 |
| 52321 | 12.70 | 4.13 | 3.45 | 4.05 | 2.99 | 2.75 | 1.94 | 5.34 | 1.78 | 1.98 | 2.19 | 1.76 | 4.31 |
| 52330 | 14.78 | 6.90 | 8.79 | 2.92 | 7.00 | 2.25 | 2.63 | 33.36 | 3.02 | 3.66 | 4.35 | 2.45 | 5.59 |
| 52413 | 15.02 | 4.65 | 12.45 | 6.69 | 11.72 | 5.40 | 2.59 | 11.37 | 2.33 | 3.57 | 4.96 | 2.43 | 4.69 |
| 52422 | 10.43 | 3.43 | 5.35 | 1.44 | 2.35 | 0.81 | 0.92 | 8.95 | 0.89 | 1.07 | 1.35 | 1.14 | 3.86 |
| 52431 | 7.85 | 3.14 | 4.56 | 1.66 | 2.45 | 1.51 | 1.18 | 6.36 | 1.15 | 1.27 | 1.45 | 1.36 | 4.14 |
| 52440 | 7.45 | 2.81 | 9.91 | 1.63 | 1.85 | 0.88 | 0.73 | 8.93 | 1.21 | 1.84 | 2.46 | 0.74 | 2.70 |
| 52450 | 16.24 | 9.24 | 9.47 | 4.50 | 6.66 | 3.16 | 2.71 | 29.21 | 2.62 | 2.66 | 2.73 | 2.38 | 9.51 |
| 52460 | 4.53 | 2.12 | 4.67 | 2.13 | 2.06 | 1.70 | 1.75 | 2.28 | 1.76 | 1.82 | 1.89 | 1.98 | 2.14 |
| 52470 | 7.12 | 3.03 | 5.03 | 2.75 | 2.18 | 2.48 | 1.75 | 6.68 | 1.65 | 1.72 | 1.82 | 1.49 | 2.70 |
| 52485 | 11.64 | 5.62 | 7.20 | 2.53 | 3.10 | 2.05 | 1.63 | 5.92 | 1.20 | 1.19 | 1.37 | 1.89 | 6.30 |
| 52491 | 5.68 | 1.43 | 4.06 | 1.43 | 1.42 | 0.51 | 0.48 | 5.40 | 0.61 | 0.91 | 1.20 | 0.59 | 1.40 |
| 52500 | 24.90 | 9.68 | 15.12 | 8.17 | 11.31 | 1.39 | 2.11 | 12.30 | 2.26 | 2.43 | 2.60 | 2.48 | 13.55 |
| 52610 | 16.35 | 7.84 | 24.05 | 4.73 | 5.98 | 5.38 | 4.97 | 32.39 | 4.63 | 4.36 | 4.14 | 5.27 | 5.03 |
| 52620 | 4.71 | 2.43 | 4.81 | 3.32 | 3.57 | 1.09 | 0.86 | 4.76 | 0.91 | 0.93 | 0.95 | 0.74 | 4.77 |
| 52630 | 9.49 | 2.26 | 2.68 | 1.67 | 2.29 | 0.71 | 1.47 | 2.31 | 1.49 | 1.45 | 1.42 | 1.83 | 2.20 |
| 52700 | 6.94 | 2.42 | 7.21 | 3.02 | 4.51 | 2.46 | 2.03 | 4.99 | 1.62 | 1.53 | 1.60 | 2.29 | 2.23 |
| median (rrmse) | 8.67 | 3.28 | 4.92 | 2.94 | 2.61 | 1.59 | 1.61 | 5.66 | **1.56** | 1.60 | 1.68 | 1.68 | 3.55 |
| mean (rrmse) | 9.86 | 3.78 | 6.92 | 3.32 | 3.99 | 1.95 | 1.67 | 9.48 | **1.64** | 1.81 | 2.02 | 1.71 | 4.16 |

The consistency of the pseudo-EBLUP brought about by using the survey weights helps to reduce the bias in some industries, but conversely increases it in others. The average effect is to increase the rrmse over the robust synthetic estimator, so it seems that adding the sampling weights is not sufficient to deal completely with the differences in business sizes.

The naïve M-quantile approach reduces the average rrmse still further than with the robust synthetic estimator, though with a similar median rrmse. It works well in this dataset, even though it is not a consistent estimator in general, producing estimates with good properties in almost all industries. There is a single larger rrmse for industry 52610, which is consistently challenging for all the approaches considered. This seems to be because 52610 includes one of the businesses with the largest Cook's distances (Fig. 2), and it is in a completely enumerated stratum in the SBS design, so this extreme outlier is included in *every* replicate sample. Introducing the bias correction to the M-quantile estimator to introduce consistency does not have the expected effect. It reduces the bias in many industries, but also substantially increases the bias in a smaller number of industry estimates, with the result that the average rrmse over industries is also substantially increased, to an unacceptable level.

The weighted versions of the naïve and bias-adjusted M-quantile estimators should both be design-consistent, and indeed we find only small biases for both, though the average relative bias for the weighted naïve M-quantile is larger than for its unweighted version. The bias of the weighted bias-adjusted M-quantile estimator is particularly low, but the compensatory increase in the variance gives a large average rrmse which means it is not satisfactory as a small area estimator in this case.

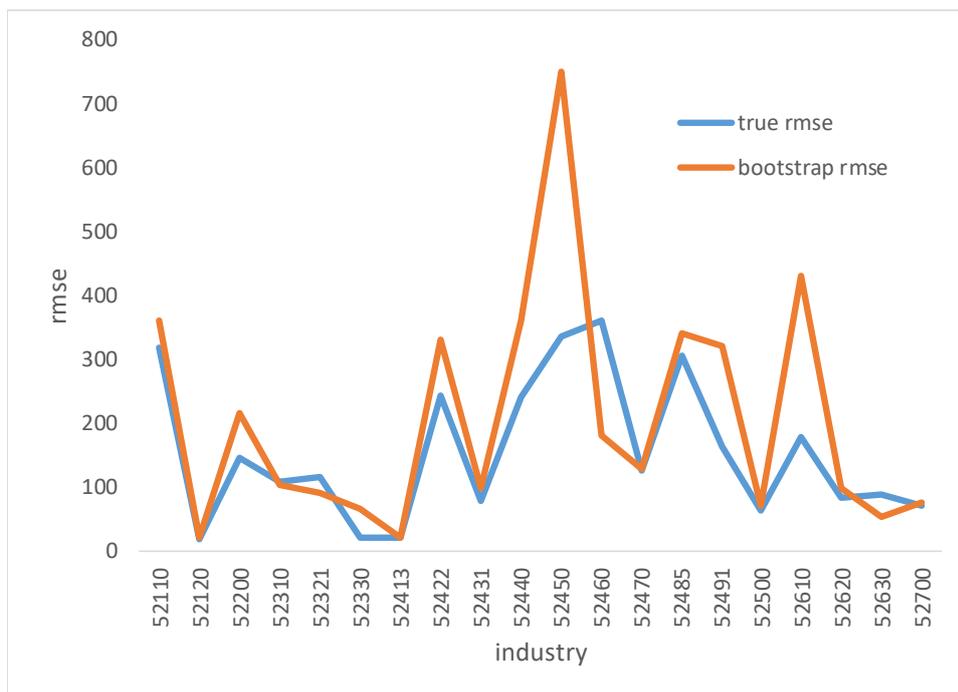

Fig. 3: The empirical true mse and mean of the bootstrap mse's for the M-quantile robustly bias-adjusted estimator.

Finally the (unweighted, predictive) robustly bias-adjusted M-quantile estimators have similar properties to the (unweighted) naïve M-quantile estimator; the estimator with $b_\phi$ = 1 is the best of the values considered, which is odd because $b_\phi < b_\psi$, contradicting expectations from theory; we return to this in section 5.1. The robustly bias-adjusted estimator with $b_\phi$ = 1 generally causes reductions in the relative biases of the industries which have the largest relative biases with the naïve

M-quantile estimator; industries with small relative biases may see an increase. But the offsetting changes in the variance from the bias correction term mean that some rrmse's go up and others down; the net effect is a marginal improvement from the best of the alternative estimators. But it is debatable whether the additional complications from this machinery would be worthwhile in this case.

4.1 Estimating the mean squared errors of the robustly bias-adjusted M-quantile estimator

Mean squared error (mse) estimation is an important part of small area estimation. In this paper we consider the estimation of the mse of the predictive M-quantile robustly bias-adjusted estimator. It is possible to construct a linearisation-type estimator of the mean-squared error for some estimators which can be written as a weighted sum of the observed values (Tzavidis et al. 2010). But for the approaches considered here there is in general no linearisation estimator, and we must fall back on a bootstrap estimator of the rmse (also presented in Tzavidis et al. 2010). Since we have the simulations from the known population, we can evaluate the bootstrap rmse estimator based on single samples with the true rmse evaluated over the 500 selected samples. The results of this comparison are presented in Table 5. With the exception of few industries, the estimated mse tracks the empirical mse fairly well as illustrated in Figure 3.

**Table 5:** Mean of the bootstrap estimates of the rmse for the robustly bias-adjusted M-quantile estimator with $b_\phi = 1$, and the true (empirical) rmse calculated over the simulations. The corresponding rrmse's are also shown

| Industry | True rmse | Mean estimated rmse | True rrmse (%) | Mean estimated rrmse (%) |
|---|---|---|---|---|
| 52110 | 317.80 | 360.72 | 0.80 | 0.91 |
| 52120 | 18.35 | 20.33 | 1.62 | 1.80 |
| 52200 | 145.28 | 215.84 | 0.44 | 0.65 |
| 52310 | 107.20 | 102.54 | 0.74 | 0.70 |
| 52321 | 114.75 | 89.91 | 1.78 | 1.40 |
| 52330 | 21.47 | 65.67 | 3.02 | 9.24 |
| 52413 | 19.72 | 20.89 | 2.33 | 2.46 |
| 52422 | 243.39 | 329.97 | 0.89 | 1.21 |
| 52431 | 77.58 | 98.76 | 1.15 | 1.47 |
| 52440 | 239.80 | 359.69 | 1.21 | 1.81 |
| 52450 | 336.25 | 749.83 | 2.62 | 5.84 |
| 52460 | 360.98 | 180.97 | 1.76 | 0.88 |
| 52470 | 125.61 | 126.97 | 1.65 | 1.67 |
| 52485 | 306.26 | 339.63 | 1.20 | 1.33 |
| 52491 | 163.40 | 319.76 | 0.61 | 1.18 |
| 52500 | 63.07 | 71.05 | 2.26 | 2.55 |
| 52610 | 179.05 | 429.75 | 4.63 | 11.12 |
| 52620 | 82.96 | 97.71 | 0.91 | 1.07 |
| 52630 | 88.78 | 54.26 | 1.49 | 0.91 |
| 52700 | 70.36 | 74.92 | 1.62 | 1.73 |
| Median | 120.18 | 114.75 | 1.56 | 1.43 |
| Mean | 154.10 | 205.46 | 1.64 | 2.50 |

## 5. Sensitivity

We have examined a wide range of estimators in section 4 to assess which have the best repeated sampling properties for use with our example business survey dataset. This provides evidence of the most appropriate methods, and how these vary as the properties (particularly the presence of outliers) of the population information differ in different small areas. In addition to the guidance provided by such an investigation, the series of simulations also allows us to assess the sensitivity of the small area estimates to some further method and estimator properties, which we describe in this section.

5.1 Impact of $b_\phi$ in the robust predictive estimators

In a practical application of small area estimation in a business survey there would not be any objective way to set the tuning constant $b_\phi$ in the bias adjustment (for the robustly bias-adjusted REBLUP or M-quantile estimator). We therefore explored the sensitivity of the rmse to the range of values in more detail. We examine $b_\phi$ from 0.25 to 3 in steps of 0.25, using the same simulated samples as for the main results in section 4 and the relative rmses for some selected industries are shown in Table 6 (Table S9 in the supplementary material gives results for all industries, from which the median and mean are calculated).

**Table 6:** Relative rmse of the robustly bias-adjusted M-quantile estimator under repeated sampling over values of $b_\phi$ from 0.25 to 3 in steps of 0.25 for selected industries, and the median and mean rrmse over all industries. (Results for all industries are given in the supplementary material Table S9.)

| $b_\phi$ | 52110 | 52321 | 52330 | 52413 | 52630 | median | mean |
|---|---|---|---|---|---|---|---|
| 0.25 | 0.84 | 1.81 | **2.70** | 2.30 | 1.47 | 1.55 | 1.63 |
| 0.50 | 0.81 | 1.75 | 2.78 | **2.14** | 1.48 | 1.55 | **1.61** |
| 0.75 | **0.80** | 1.75 | 2.90 | 2.17 | 1.49 | 1.55 | 1.62 |
| 1.00 | 0.80 | 1.78 | 3.02 | 2.33 | 1.49 | 1.56 | 1.64 |
| 1.25 | 0.81 | 1.83 | 3.16 | 2.57 | 1.48 | 1.53 | 1.67 |
| 1.50 | 0.83 | 1.88 | 3.32 | 2.88 | 1.47 | 1.53 | 1.71 |
| 1.75 | 0.84 | 1.93 | 3.49 | 3.22 | 1.46 | 1.59 | 1.76 |
| 2.00 | 0.86 | 1.98 | 3.66 | 3.57 | 1.45 | 1.60 | 1.81 |
| 2.25 | 0.88 | 2.03 | 3.83 | 3.93 | 1.44 | 1.62 | 1.86 |
| 2.50 | 0.91 | 2.09 | 4.00 | 4.28 | 1.43 | 1.64 | 1.92 |
| 2.75 | 0.93 | 2.14 | 4.17 | 4.63 | 1.42 | 1.66 | 1.97 |
| 3.00 | 0.96 | 2.19 | 4.35 | 4.96 | **1.42** | 1.68 | 2.02 |
| max - min | 0.16 | 0.45 | 1.65 | 2.82 | 0.07 | 0.15 | 0.41 |

We hope for a minimum of the rrmse over the range of values we consider, and in our example there is a minimum in the mean of the relative rmse at $b_\phi$ = 0.5 (Fig. 4a). However, there is a range of patterns in the individual industries (Tables 6 and S9 and Fig. S2). Half of the industries have the same pattern as Fig. 4(a) with a minimum at values of $b_\phi$ from 0.5 to 1.75. Seven are monotonic increasing over the range of $b_\phi$ values considered, and one is monotonic decreasing. Two industries show a *maximum* of the rrmse (eg industry 52630, Fig. 4b), which is an unexpected pattern. Some uncertainty may be expected in the rrmse estimates for the smallest values of $b_\phi$, because the estimator is unstable when the Huber function affects many values. But the patterns in general seem to be quite stable and interpretable. It is not straightforward to pick a single value for use in all industries in this dataset, but perhaps $b_\phi = 0.75$ would be the best compromise giving reasonable bias adjustment in all industries (and indeed only values of $b_\phi < 1.25$ are better than the naïve M-quantile estimator). In most industries (eg 52110 in Table 6) the differences in the rrmse over the range of values of $b_\phi$ is small. The largest difference, in industry 52413, is almost 3% in the rrmse, which is sufficiently large that it may affect the inference on the estimates. Therefore in this example the choice of $b_\phi$ does seem to be important. We therefore suggest that the sensitivity of the estimates to different values of $b_\phi$ should be investigated when using the robustly bias-adjusted M-quantile estimator. Of course in other domains or datasets $b_\phi$ may have other impacts, and further experimentation with other business survey data and a range of estimators would be valuable.

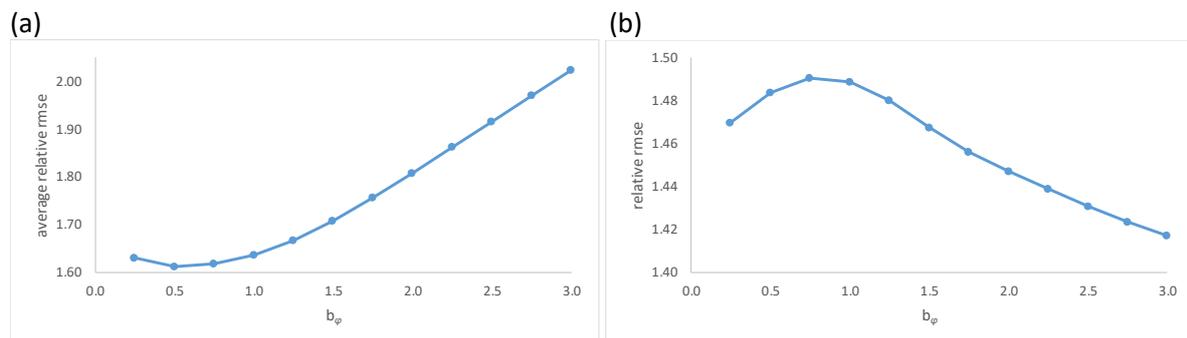

Fig. 4: The change in (a) average (over industries) relative rmse and (b) relative rmse for industry 52630, both as $b_\phi$ varies in the robustly bias-adjusted M-quantile estimator.

It is interesting that values of $b_\phi < b_\psi$, which are precluded by Tzavidis et al. (2010) – presumably because they imply *more* bias in the bias-variance trade-off in the adjustment and therefore do not in concept make a bias adjustment – seem to work well. In general, small values of $b_\phi$ give the lowest rrmses. An alternative interpretation could be that bias is not the main influence in this optimisation, and that variance may be more important.

5.2 Sensitivity to extreme population outliers

The tax data that we have used as a population contain some very extreme outliers, and this is clearly a situation in which robust methods should be effective. But we are also interested in their performance with fewer, less extreme outliers. For this purpose we produced a version of the population where the points with the highest leverage in the working model formed from the fixed part of (4) were removed, the working model was refitted, and the influence diagnostics examined. If necessary, further high leverage points were removed. This led to a reduced population where 29 of the most extreme observations were removed. The reduced population has a less skewed distribution of residuals, but the estimation of small area (industry) totals is still affected by the small sample sizes, so the same arguments for the use of small area estimators apply.

(Note that this is not proposed as a way to deal with outliers – as this process would have two significant challenges, in when to stop, and how to deal with the removed observations. But the reduced population does allow us to examine the properties of the considered estimators in a related dataset with fewer very extreme observations.)

The results of using the reduced population are presented in the supplementary material Tables S3-S5. Essentially they present the same picture as the results from the full population. The difference between the two best estimators, the naïve M-quantile estimator and the robustly bias-adjusted M-quantile estimator, is even slightly smaller. So our conclusions are not affected when the population of interest has fewer (and less extreme) outliers. Of course we are not able from this study to examine real data containing more (and more extreme) outliers, though it would be possible to introduce some artificial contaminating values and assess the sensitivity to these. We leave this for further work, although perhaps repeating the study with different populations would be a higher priority.

### 5.3 Sensitivity to model choice

Equations (4) and (5) present the two models which we have considered for this dataset, and a summary of the estimator properties using the two models (with the original, complete dataset) is shown in Table 7. The conclusions from the two models are substantively similar, though there are some small differences, which are biggest in the relative bias of the EBLUP and naïve M-quantile estimators. The full results from the simulation with model (5) are shown in tables S6-S8 in the supplementary material.

**Table 7:** Comparison of RRMSE and bias properties for the same forms of estimator and the same data, but using model (4) or (5). The GREG is based on the SBS model variables, and the HT estimator has no predictors, so neither is linked to models (4) and (5). The methods with the best properties are highlighted in bold.

|  | mean rb | | mean rrmse | |
|---|---|---|---|---|
|  | (4) | (5) | (4) | (5) |
| Direct (HT) | **0.02** | | 9.86 | |
| Direct (GREG) | 0.49 | | 3.78 | |
| EBLUP | -0.29 | -1.75 | 6.92 | 7.55 |
| EBLUP (var = f(SC)) | 0.71 | 0.62 | 3.32 | 3.08 |
| pseudo-EBLUP | 0.98 | 0.97 | 3.99 | 3.98 |
| Robust synthetic | 0.20 | 0.26 | 1.95 | 1.95 |
| M-quantile naïve ( $b_\psi = 1.345$ ) | -0.19 | **0.01** | 1.67 | 1.60 |
| M-quantile bias-adjusted ( $b_\psi = 1.345$ ) | -0.33 | -0.41 | 9.48 | 9.49 |
| M-quantile robustly bias-adjusted ( $b_\psi = 1.345$ , $b_\phi$ = 1) | -0.25 | -0.11 | **1.64** | **1.58** |
| M-quantile robustly bias-adjusted ( $b_\psi = 1.345$ , $b_\phi$ = 2) | -0.30 | -0.20 | 1.81 | 1.76 |
| M-quantile robustly bias-adjusted ( $b_\psi = 1.345$ , $b_\phi$ = 3) | -0.37 | -0.30 | 2.02 | 1.98 |
| weighted M-quantile naïve ( $b_\psi = 1.345$ ) | -0.47 | -0.41 | 1.71 | 1.67 |
| weighted M-quantile bias-adjusted ( $b_\psi = 1.345$ ) | **0.02** | **0.02** | 4.16 | 4.17 |

### 5.4 Weights

The pseudo-EBLUP and weighted M-quantile methods use the sampling weights, which means that they give design-consistent estimators in small areas as the sample size approaches the population size. This provides some extra protection against model misspecification bias, but the use of the weights increases the variance. These effects are not obvious among the range of estimators that we consider – although the bias should be reduced, there are unweighted estimators that have small bias too. The variance does however seem to be larger, though the weighted naïve M-quantile estimator is competitive in terms of rrmse in our example data. But the best-performing estimators in our example are unweighted. Applications of these estimators to further examples are needed to assess whether this is data specific or whether a more general result can be deduced.

We also considered a variation on the calculation of $\bar{q}$ in the weighted versions of (11) and (13) to use the survey weights, replacing $\bar{q}$ by $\tilde{q}_{ind} = \dfrac{\sum_{i \in ind} w_i q_i}{\sum_{i \in ind} w_i}$; this seems to use the weights twice, and the results (not shown) are not better than the original versions of the estimators, so we do not consider this approach further.

5.5    Pure prediction or out-of-sample prediction

Several of the estimators (see (6), (9), (10), (11), (15)) are of the form $\hat{y}_{ind} = \sum_{i \in ind \cap s} y_i + \sum_{i \in ind \cap r} \hat{y}_i$ for various definitions of the predictor $\hat{y}_i$, and producing estimates in this way seems to be standard practice. There is an alternative approach however, in which even the sample values are predicted, that is $\hat{y}_{ind} = \sum_{i \in ind} \hat{y}_i$ which is sometimes used, particularly when the sample is a small proportion of the population. We have used both approaches with the naïve M-quantile estimator (11) in unweighted and weighted versions. In both cases the average rrmse is lower for the first formulation, and the effect is particularly noticeable in the presence of a large outlier (though when there is no outlier there are cases of pure prediction industry estimates with lower rrmse's than the out-of-sample predictions). So (unsurprisingly) the pure prediction is rarely better than the real observation, and we recommend that for business surveys in particular the real data are always used directly when the estimator has this form.

## 6. Discussion

In this paper we have illustrated the effects of robust unit-level models as the basis for small-area estimation for business surveys with variables with skewed distributions; this contrasts with the adaptation of area-level models to business surveys in Ferrante & Pacei 2017 and Fabrizi *et al*. 2018. M-quantile methods which deal with outlying observations provide a strategy for small domain estimation which offers much reduced rmses over direct estimation when domains are small, and better properties than other robust small area methods in this example. The robustly bias-adjusted M-quantile estimator with $b_\phi$ = 0.5 produces the best overall results, but the unweighted and weighted naïve M-quantile estimators have similar performance, and with these latter methods there are fewer complications from the need to choose an appropriate tuning parameter. The relative rmse's for industry-level small area estimates in Table 4 range from 0.44 to 4.63%, which makes estimates much more usable than the corresponding (direct) Horvitz-Thompson estimates which have relative rmse's from 2.98 to 24.90%, and even considerably better than the GREG which uses the strong auxiliary information in design-based estimation, where the industry rrmse's range from 1.01 to 9.68%.

After this research project was finished, Statistics Netherlands decided to change the design of the SBS, for reasons not connected with this research. As a consequence, there is less need for these types of small area estimation methods, and the approaches developed in this paper have not been implemented. Nevertheless the research is useful, as similar problems occur in other statistics and in other countries.

In the M-quantile methods of section (3.3) we used $b_\psi$ = 1.345 as the tuning constant in (8), which is optimal for normally distributed errors under certain conditions (Holland & Welsch 1977, Dawber *et al*. in prep). However, we know that at least some of the difficulty with business surveys is that the errors are not normally distributed. Dawber *et al*. (in prep) propose numerical methods to deduce optimal values of the tuning constant for a range of distributions; these vary according to the value of $q$.

The robustly bias-adjusted M-quantile estimator is the best in our example, but shows only small gains over the naïve M-quantile estimator, and the result depends on the value chosen for the tuning constant $b_\phi$ in the bias adjustment. In practice there would be no simulation results to indicate the best value for the tuning constant, and in order to have an evidence base for this choice it would be helpful to have further studies with known business survey populations to examine how much this value varies. Until this is available we suggest assessing the sensitivity of estimates to the choice of this parameter. The examination of the average relative rmse with varying values of $b_\phi$ shows that the rmse's are sensitive to this parameter in some cases, and could have an impact on inferences in the most affected estimates. The choice of $b_\phi < b_\psi$ is best in our example, a situation not expected to be effective according to the way this theory is developed (Tzavidis *et al*. 2010).

There are some outstanding issues that would benefit from further research. We have used tax data in different years as predictor and outcome variables, but it would be interesting to see whether the results are maintained in a situation with survey data, particularly where the survey response is a different variable from the tax data which act as a predictor. It is however typical in business surveys for there to be strong predictors for size-related outcome variables, derived from administrative data and often available operationally from the business register. For variables where the predictors are not so strong, small area estimation approaches will likely have a smaller impact on the rmse, and it seems likely that robust methods will only be effective if the outliers are the main causes of lack of fit. Further investigation of these approaches with different types of variables would therefore be valuable to gauge the limits of their usefulness.

We used small area estimation in isolation, to better understand its properties. In practice, however, we would like to enforce consistency with the main SBS outputs. Benchmarking to the direct estimators at higher levels of aggregation where the rmse is small could achieve this (Pfeffermann et al. 2014), and might also have beneficial impacts on the bias and rmse of the small area estimators. Alternatively, if the aggregated small area estimates are more accurate than the direct estimate of the population total, then benchmarking may introduce too much of the random variation in the direct estimation. We leave this for further investigation.

# Supplementary material

for "Robust estimation for small domains in business surveys" by Paul A. Smith, Chiara Bocci, Nikos Tzavidis, Sabine Krieg, Marc J.E. Smeets

## S1: Additional tables

Table S1 shows the AIC, BIC and log-likelihood values which support modelling decisions described in the text.

Table S1: Model fit statistics for EBLUP method (6) and different models.

| Model | AIC | BIC | logL |
|---|---|---|---|
| EBLUP (6) & model (4) | 29,552.24 | 29,617.55 | -14,766.12 |
| EBLUP (6) & model (5) | 29,655.50 | 29,707.75 | -14,819.75 |
| EBLUP (6) & model (4), $\mathbf{e}_{sc} \sim N(\mathbf{0}, \mathbf{\Sigma}_{e,sc})$ | 19,501.94 | 19,593.36 | -9,736.97 |
| EBLUP (6) & model (4), $\mathbf{e} = \mathbf{wp}^T \mathbf{\varepsilon}$ with $\mathbf{\varepsilon} \sim N(\mathbf{0}, \mathbf{\Sigma}_\varepsilon)$ | 20,990.68 | 21,055.99 | -10,485.34 |
| EBLUP (6) & model (4), $\mathbf{e} = (\mathbf{wp}^2)^T \mathbf{\varepsilon}$ with $\mathbf{\varepsilon} \sim N(\mathbf{0}, \mathbf{\Sigma}_\varepsilon)$ | 19,113.68 | 19,178.98 | -9,546.84 |

Table S2 restates the summary results from tables 2 and 3 of the main paper in a form which is easier for comparative reading.

Table S2: Summary information on relative bias (from Table 2) and relative rmse (from Table 3) for the models listed in Table 1. The best-performing models are indicated in bold.

| | mean rb | mean abs(rb) | mean rrmse |
|---|---|---|---|
| Direct (HT) | **0.02** | 0.34 | 9.86 |
| Direct (GREG) | 0.49 | 0.93 | 3.78 |
| EBLUP | -0.29 | 3.81 | 6.92 |
| EBLUP (var = f(SC)) | 0.71 | 1.60 | 3.32 |
| pseudo-EBLUP | 0.98 | 1.68 | 3.99 |
| Robust synthetic | 0.20 | 1.75 | 1.95 |
| M-quantile naïve ($b_\psi = 1.345$) | -0.19 | 1.25 | 1.67 |
| M-quantile bias-adjusted ($b_\psi = 1.345$) | -0.33 | 5.71 | 9.48 |
| M-quantile robustly bias-adjusted ($b_\psi = 1.345$, $b_\phi = 1$) | -0.25 | 0.99 | **1.64** |
| M-quantile robustly bias-adjusted ($b_\psi = 1.345$, $b_\phi = 2$) | -0.30 | 1.02 | 1.81 |
| M-quantile robustly bias-adjusted ($b_\psi = 1.345$, $b_\phi = 3$) | -0.37 | 1.10 | 2.02 |
| weighted M-quantile naïve ($b_\psi = 1.345$) | -0.47 | 1.25 | 1.71 |
| weighted M-quantile bias-adjusted ($b_\psi = 1.345$) | **0.02** | **0.16** | 4.16 |

## S2: Reduced population

This section gives results from a smaller population with the most extreme outliers removed, as described in section 5.2 of the main paper. Fig. S1 shows the outlier diagnostics for the reduced population relative to the working model, which is the fixed part of (4).

Fig S1: (a) Residuals and (b) Cook's distance from reduced population.

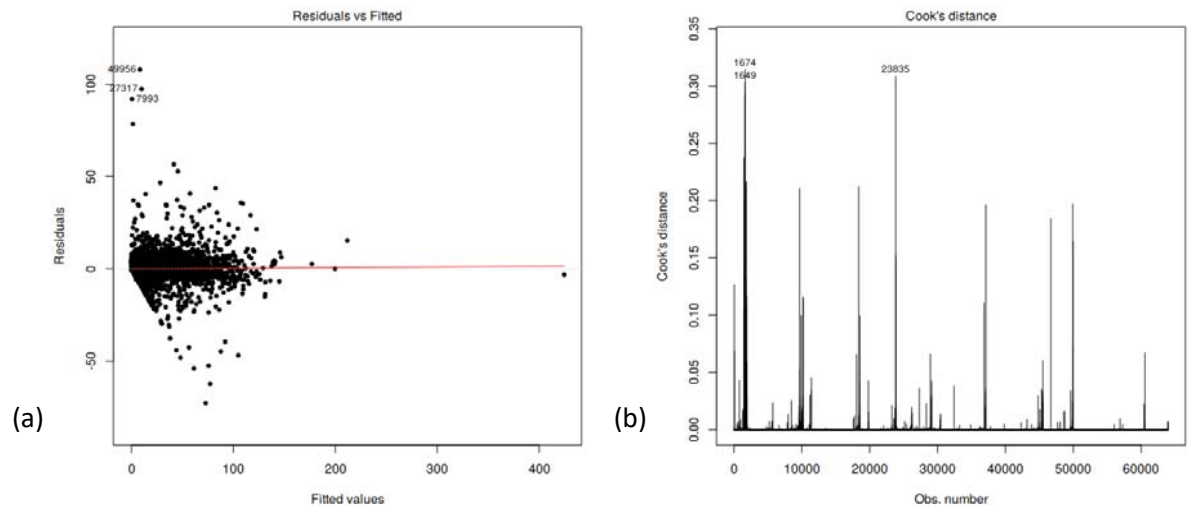

Table S3: Industry-specific relative bias (rb) (%) of small area point estimators of total *tto* in the reduced population. The best-performing models are indicated in bold in the summary lines.

| Industry | Direct | | EBLUP | EBLUP | pseudo-EBLUP | robust synthetic | M-quantile naïve | M-quantile bias-adjusted | M-quantile robustly bias-adjusted | | | M-quantile naïve | M-quantile bias-adjusted |
|---|---|---|---|---|---|---|---|---|---|---|---|---|---|
| | HT | GREG | | | | | | | unweighted | | | weighted | |
| | | | | var = f(sc) | | | | | $b_\phi = 1$ | $b_\phi = 2$ | $b_\phi = 3$ | | |
| 52110 | -0.09 | 0.23 | 0.34 | 0.52 | 0.24 | 0.49 | 0.15 | 0.41 | -0.10 | -0.10 | -0.04 | -0.10 | 0.03 |
| 52120 | -0.67 | -1.05 | 1.64 | 1.71 | 2.13 | 1.26 | 0.50 | 0.23 | 0.25 | 0.01 | -0.19 | 0.29 | -0.10 |
| 52200 | 0.13 | -0.57 | 1.15 | 1.77 | 1.07 | 1.57 | 0.62 | 0.83 | 0.32 | 0.19 | 0.15 | 0.37 | -0.08 |
| 52310 | 0.27 | -0.25 | 0.70 | -0.18 | -0.32 | 0.02 | 0.52 | 0.55 | 0.67 | 0.67 | 0.62 | 0.36 | 0.01 |
| 52321 | -0.32 | 1.67 | 2.70 | 2.51 | 2.95 | 2.83 | 1.80 | 0.24 | 1.33 | 1.28 | 1.28 | 1.53 | -0.27 |
| 52330 | 0.26 | -0.76 | -0.98 | 1.57 | 5.99 | 1.23 | 0.84 | -31.72 | 0.94 | 0.58 | 0.01 | 0.50 | -0.50 |
| 52413 | 0.48 | 0.11 | 1.39 | 5.38 | 6.51 | 5.34 | 2.24 | -11.32 | 0.02 | -1.80 | -3.33 | 1.96 | -0.04 |
| 52422 | 0.05 | 0.70 | 1.50 | 0.16 | -0.11 | 0.01 | 0.00 | 7.44 | 0.30 | 0.65 | 0.86 | -0.30 | 0.03 |
| 52431 | -0.04 | -0.31 | -1.31 | -1.55 | -1.75 | -2.09 | -1.47 | 0.86 | -1.19 | -0.82 | -0.58 | -1.69 | 0.06 |
| 52440 | 0.12 | -0.06 | -1.47 | 0.45 | 0.69 | 0.73 | 0.15 | -5.17 | -0.63 | -1.28 | -1.88 | -0.21 | -0.14 |
| 52450 | 0.65 | -0.14 | 0.14 | 0.58 | 0.92 | 1.44 | 0.77 | -1.21 | 0.53 | 0.50 | 0.47 | 0.36 | 0.06 |
| 52460 | 0.20 | -0.26 | -1.81 | -1.13 | -1.13 | -1.45 | -1.51 | -1.60 | -1.51 | -1.56 | -1.61 | -1.77 | 0.01 |
| 52470 | 0.44 | 0.72 | 2.02 | 1.72 | 1.85 | 2.12 | 1.32 | 1.52 | 1.07 | 1.01 | 0.99 | 0.98 | 0.11 |
| 52485 | 0.31 | 0.11 | -0.61 | -1.15 | -1.42 | -1.52 | -0.99 | 1.81 | -0.33 | -0.03 | -0.05 | -1.32 | -0.12 |
| 52491 | 0.09 | 1.11 | 0.96 | 0.24 | -0.28 | -0.60 | -0.22 | 3.08 | 0.19 | 0.56 | 0.86 | -0.55 | 0.04 |
| 52500 | 1.25 | -0.30 | -2.83 | 1.43 | -0.83 | -1.33 | -1.43 | 5.53 | -1.32 | -1.19 | -1.06 | -2.08 | 0.40 |
| 52610 | 0.02 | 5.64 | -0.85 | -4.94 | -6.74 | -7.26 | -6.25 | 8.63 | -5.76 | -5.35 | -5.01 | -6.82 | -0.24 |
| 52620 | 0.22 | 0.19 | -1.38 | 1.08 | -0.02 | -0.31 | -0.48 | 0.24 | -0.49 | -0.54 | -0.59 | -0.85 | 0.12 |
| 52630 | 0.07 | -0.21 | -1.10 | -0.20 | -0.38 | -0.44 | -0.96 | -0.59 | -1.02 | -1.00 | -0.98 | -1.42 | 0.15 |
| 52700 | 0.05 | 0.59 | -3.20 | -0.27 | -1.88 | -2.44 | -1.58 | 1.81 | -1.16 | -0.95 | -0.89 | -2.04 | 0.01 |
| median (rb) | 0.12 | **0.02** | -0.24 | 0.48 | -0.06 | **0.01** | 0.07 | 0.48 | -0.04 | -0.06 | -0.12 | -0.25 | **0.01** |
| mean (rb) | 0.17 | 0.36 | -0.15 | 0.49 | 0.37 | **-0.02** | -0.30 | -0.92 | -0.39 | -0.46 | -0.55 | -0.64 | **-0.02** |
| mean abs(rb) | 0.28 | 0.75 | 1.40 | 1.43 | 1.86 | 1.72 | 1.19 | 4.24 | 0.96 | 1.00 | 1.07 | 1.27 | **0.13** |

Table S4: Industry-specific relative rmse (%) of small area point estimators of the total *tto* in the reduced population. The best-performing models are indicated in bold in the summary lines.

| Industry | Direct | | EBLUP | EBLUP | pseudo-EBLUP | robust synthetic | M-quantile naïve | M-quantile bias-adjusted | M-quantile robustly bias-adjusted | | | M-quantile naïve | M-quantile bias-adjusted |
|---|---|---|---|---|---|---|---|---|---|---|---|---|---|
| | HT | GREG | | | | | | | unweighted | | | weighted | |
| | | | | var = f(sc) | | | | | $b_\phi = 1$ | $b_\phi = 2$ | $b_\phi = 3$ | | |
| 52110 | 3.36 | 1.38 | 1.37 | 1.00 | 1.04 | 0.63 | 0.55 | 2.18 | 0.61 | 0.69 | 0.77 | 0.61 | 1.71 |
| 52120 | 10.51 | 3.20 | 2.23 | 2.48 | 2.29 | 1.75 | 1.61 | 3.31 | 1.60 | 1.65 | 1.72 | 1.59 | 3.27 |
| 52200 | 3.95 | 1.87 | 1.77 | 2.13 | 1.63 | 1.59 | 0.72 | 2.02 | 0.52 | 0.50 | 0.54 | 0.54 | 1.76 |
| 52310 | 2.95 | 1.00 | 0.93 | 0.89 | 0.90 | 0.33 | 0.66 | 0.88 | 0.78 | 0.79 | 0.76 | 0.55 | 0.94 |
| 52321 | 14.36 | 3.23 | 3.10 | 3.25 | 3.11 | 2.85 | 2.00 | 5.00 | 1.77 | 1.94 | 2.14 | 1.78 | 4.28 |
| 52330 | 14.20 | 5.69 | 4.04 | 2.51 | 6.07 | 2.09 | 2.37 | 35.07 | 2.75 | 3.32 | 4.04 | 2.26 | 5.47 |
| 52413 | 14.49 | 4.15 | 4.02 | 5.98 | 6.75 | 5.36 | 2.60 | 12.33 | 2.15 | 3.36 | 4.80 | 2.37 | 4.26 |
| 52422 | 9.38 | 1.91 | 2.60 | 0.93 | 0.83 | 0.26 | 0.48 | 8.67 | 0.83 | 1.23 | 1.56 | 0.61 | 1.94 |
| 52431 | 7.76 | 2.11 | 1.80 | 1.79 | 1.92 | 2.12 | 1.65 | 4.31 | 1.52 | 1.39 | 1.45 | 1.87 | 2.53 |
| 52440 | 7.16 | 2.26 | 3.28 | 1.29 | 1.16 | 0.79 | 0.60 | 7.69 | 1.17 | 1.83 | 2.45 | 0.69 | 2.47 |
| 52450 | 7.40 | 2.26 | 1.88 | 1.55 | 1.50 | 1.49 | 1.02 | 3.84 | 1.09 | 1.29 | 1.46 | 0.81 | 3.09 |
| 52460 | 4.41 | 1.87 | 2.16 | 1.76 | 1.52 | 1.49 | 1.57 | 2.36 | 1.61 | 1.70 | 1.78 | 1.83 | 2.07 |
| 52470 | 7.15 | 2.23 | 2.35 | 2.29 | 2.10 | 2.15 | 1.50 | 3.70 | 1.41 | 1.52 | 1.65 | 1.24 | 2.75 |
| 52485 | 6.93 | 2.35 | 2.48 | 1.70 | 1.79 | 1.55 | 1.14 | 5.38 | 0.93 | 1.14 | 1.40 | 1.46 | 2.48 |
| 52491 | 3.74 | 1.48 | 1.64 | 0.90 | 0.78 | 0.65 | 0.40 | 3.50 | 0.43 | 0.72 | 1.00 | 0.66 | 1.21 |
| 52500 | 23.63 | 6.38 | 4.80 | 6.05 | 3.52 | 1.45 | 1.81 | 11.94 | 1.96 | 2.17 | 2.38 | 2.38 | 13.20 |
| 52610 | 16.88 | 12.06 | 8.64 | 6.24 | 8.41 | 7.64 | 6.72 | 14.36 | 6.29 | 5.92 | 5.63 | 7.27 | 6.79 |
| 52620 | 4.52 | 1.20 | 2.10 | 1.80 | 1.31 | 0.45 | 0.66 | 2.10 | 0.71 | 0.79 | 0.87 | 0.98 | 1.51 |
| 52630 | 9.41 | 2.55 | 1.83 | 1.18 | 1.56 | 0.64 | 1.11 | 2.27 | 1.18 | 1.18 | 1.18 | 1.57 | 2.27 |
| 52700 | 7.61 | 2.57 | 3.99 | 1.79 | 2.48 | 2.47 | 1.73 | 4.90 | 1.43 | 1.41 | 1.53 | 2.17 | 2.20 |
| median (rrmse) | 7.51 | 2.26 | 2.29 | 1.79 | 1.71 | 1.52 | **1.32** | 4.08 | **1.30** | 1.40 | 1.54 | 1.51 | 2.48 |
| mean (rrmse) | 8.99 | 3.09 | 2.85 | 2.38 | 2.53 | 1.89 | **1.55** | 6.79 | **1.54** | 1.73 | 1.96 | 1.66 | 3.31 |

Table S5: Summary information on relative bias (from Table S3) and relative rmse (from Table S4) for the models listed in Table 1 using the reduced population described in section 5.2. The best-performing models are indicated in bold.

|  | mean rb | mean abs(rb) | mean rrmse |
|---|---|---|---|
| Direct (HT) | 0.17 | 0.28 | 8.99 |
| Direct (GREG) | 0.36 | 0.75 | 3.09 |
| EBLUP | -0.15 | 1.40 | 2.85 |
| EBLUP (var = f(SC)) | 0.49 | 1.43 | 2.38 |
| pseudo-EBLUP | 0.37 | 1.86 | 2.53 |
| Robust synthetic | -0.02 | 1.72 | 1.89 |
| M-quantile naïve ( $b_\psi$ = 1.345 ) | -0.30 | 1.19 | **1.55** |
| M-quantile bias-adjusted ( $b_\psi$ = 1.345 ) | -0.92 | 4.24 | 6.79 |
| M-quantile robustly bias-adjusted ( $b_\psi$ = 1.345 , $b_\phi$ = 1) | -0.39 | 0.96 | 1.54 |
| M-quantile robustly bias-adjusted ( $b_\psi$ = 1.345 , $b_\phi$ = 2) | -0.46 | 1.00 | 1.73 |
| M-quantile robustly bias-adjusted ( $b_\psi$ = 1.345 , $b_\phi$ = 3) | -0.55 | 1.07 | 1.96 |
| weighted M-quantile naïve ( $b_\psi$ = 1.345 ) | -0.64 | 1.27 | 1.66 |
| weighted M-quantile bias-adjusted ( $b_\psi$ = 1.345 ) | **-0.02** | **0.13** | 3.31 |

**S3: Reduced model**

This section gives results from the original, complete population but with models including the reduced set of variables in equation (5). The same range of estimators is used.

Table S6: Industry-specific relative bias (rb) (%) of small area point estimators of total *tto* in the full population using model (5). The best-performing models are indicated in bold in the summary lines.

| Industry | Direct | | EBLUP | EBLUP | pseudo-EBLUP | robust synthetic | M-quantile naïve | M-quantile bias-adjusted | M-quantile robustly bias-adjusted | | | M-quantile naïve | M-quantile bias-adjusted |
|---|---|---|---|---|---|---|---|---|---|---|---|---|---|
| | HT | GREG | | | | | | | unweighted | | | weighted | |
| | | | | var = f(sc) | | | | | $b_\phi = 1$ | $b_\phi = 2$ | $b_\phi = 3$ | | |
| 52110 | -0.28 | 0.51 | -0.84 | 0.64 | -0.24 | 0.96 | 0.58 | 0.17 | 0.34 | 0.36 | 0.43 | 0.26 | -0.05 |
| 52120 | -0.59 | -1.16 | 1.11 | 2.26 | 1.82 | 1.20 | 0.47 | 0.08 | 0.21 | -0.03 | -0.24 | 0.23 | -0.22 |
| 52200 | 0.08 | -0.50 | 0.66 | 1.43 | 0.38 | 1.35 | 0.41 | 1.43 | 0.09 | -0.05 | -0.09 | 0.10 | 0.04 |
| 52310 | -0.16 | -0.02 | -0.16 | -0.60 | -0.81 | 0.05 | 0.52 | 0.06 | 0.67 | 0.65 | 0.59 | 0.30 | -0.02 |
| 52321 | -0.27 | 1.80 | -0.21 | 2.24 | 1.88 | 2.87 | 1.84 | 0.37 | 1.34 | 1.26 | 1.27 | 1.49 | 0.20 |
| 52330 | -0.24 | -0.76 | -9.68 | 1.48 | 5.34 | 1.45 | 1.35 | -29.81 | 1.35 | 1.02 | 0.49 | 0.83 | 0.11 |
| 52413 | 0.48 | 0.26 | -11.87 | 5.12 | 8.08 | 5.35 | 2.32 | -10.83 | 0.05 | -1.80 | -3.31 | 1.93 | 0.21 |
| 52422 | -0.48 | -0.30 | -0.04 | -0.62 | -0.52 | -0.60 | -0.61 | 7.40 | -0.34 | 0.01 | 0.24 | -0.97 | -0.22 |
| 52431 | 0.19 | 1.02 | -2.52 | -1.29 | -1.11 | -1.32 | -0.73 | -0.21 | -0.51 | -0.26 | -0.07 | -1.09 | -0.50 |
| 52440 | 0.26 | 0.18 | -8.38 | 0.53 | 0.29 | 0.93 | 0.38 | -5.95 | -0.45 | -1.14 | -1.82 | -0.05 | 0.00 |
| 52450 | 0.56 | 0.89 | -2.59 | 1.50 | 1.67 | 2.60 | 1.88 | -10.13 | 1.67 | 1.62 | 1.60 | 1.32 | 0.10 |
| 52460 | 0.29 | -0.31 | -3.80 | -1.16 | -0.70 | -1.50 | -1.57 | -1.04 | -1.59 | -1.64 | -1.67 | -1.89 | 0.11 |
| 52470 | 0.06 | 0.58 | -2.07 | 1.65 | 0.82 | 2.57 | 1.65 | -2.10 | 1.38 | 1.28 | 1.23 | 1.25 | -0.14 |
| 52485 | 0.33 | -0.53 | -4.45 | -1.49 | -0.74 | -1.87 | -1.37 | -0.09 | -0.75 | -0.48 | -0.52 | -1.77 | 0.13 |
| 52491 | -0.01 | 1.76 | 1.20 | 0.37 | 0.39 | -0.34 | 0.08 | 4.62 | 0.47 | 0.84 | 1.13 | -0.32 | 0.03 |
| 52500 | 0.90 | -0.48 | -4.60 | 1.83 | 3.18 | -1.40 | -1.30 | 5.74 | -1.18 | -1.07 | -0.94 | -2.09 | 0.65 |
| 52610 | -0.68 | 4.18 | 22.54 | -3.25 | -2.64 | -5.04 | -4.07 | 30.77 | -3.72 | -3.42 | -3.18 | -4.66 | -0.05 |
| 52620 | 0.31 | 2.56 | -2.16 | 2.26 | 1.21 | 0.94 | 0.82 | 0.35 | 0.82 | 0.76 | 0.71 | 0.43 | 0.13 |
| 52630 | -0.48 | -0.35 | -1.12 | -0.22 | 0.17 | -0.44 | -0.86 | -0.75 | -0.93 | -0.93 | -0.92 | -1.34 | -0.05 |
| 52700 | 0.12 | 0.53 | -6.01 | -0.27 | 0.99 | -2.49 | -1.60 | 1.79 | -1.12 | -0.89 | -0.83 | -2.06 | -0.09 |
| median (rb) | 0.07 | 0.22 | -2.12 | 0.59 | 0.38 | 0.49 | 0.39 | 0.07 | 0.07 | -0.04 | -0.08 | 0.03 | **0.01** |
| mean (rb) | **0.02** | 0.49 | -1.75 | 0.62 | 0.97 | 0.26 | **0.01** | -0.41 | -0.11 | -0.20 | -0.30 | -0.41 | **0.02** |
| mean abs(rb) | 0.34 | 0.93 | 4.30 | 1.51 | 1.65 | 1.76 | 1.22 | 5.68 | 0.95 | 0.98 | 1.06 | 1.22 | **0.15** |

Table S7: Industry-specific relative rmse (%) of small area point estimators of the total *tto* in the full population using model (5). The best-performing models are indicated in bold in the summary lines.

| Industry | Direct | | EBLUP | EBLUP | pseudo-EBLUP | robust synthetic | M-quantile naïve | M-quantile bias-adjusted | M-quantile robustly bias-adjusted | | | M-quantile naïve | M-quantile bias-adjusted |
|---|---|---|---|---|---|---|---|---|---|---|---|---|---|
| | HT | GREG | | | | | | | unweighted | | | weighted | |
| | | | | var = $f(sc)$ | | | | | $b_\phi = 1$ | $b_\phi = 2$ | $b_\phi = 3$ | | |
| 52110 | 3.88 | 1.72 | 2.86 | 2.22 | 2.04 | 1.07 | 0.85 | 2.64 | 0.77 | 0.85 | 0.97 | 0.72 | 2.08 |
| 52120 | 10.22 | 3.23 | 2.56 | 3.03 | 2.55 | 1.70 | 1.62 | 3.27 | 1.62 | 1.66 | 1.74 | 1.60 | 3.24 |
| 52200 | 4.33 | 2.06 | 2.37 | 2.37 | 1.76 | 1.38 | 0.56 | 2.49 | 0.43 | 0.48 | 0.53 | 0.42 | 1.86 |
| 52310 | 2.98 | 1.02 | 1.40 | 1.95 | 2.02 | 0.37 | 0.67 | 0.74 | 0.78 | 0.78 | 0.74 | 0.53 | 1.00 |
| 52321 | 12.70 | 3.32 | 3.30 | 3.75 | 2.81 | 2.90 | 2.06 | 5.34 | 1.83 | 2.00 | 2.21 | 1.76 | 4.35 |
| 52330 | 14.78 | 5.80 | 12.54 | 2.59 | 5.96 | 2.28 | 2.73 | 33.71 | 3.06 | 3.65 | 4.32 | 2.49 | 5.60 |
| 52413 | 15.02 | 4.79 | 15.30 | 6.83 | 11.73 | 5.37 | 2.71 | 11.84 | 2.28 | 3.49 | 4.90 | 2.40 | 4.70 |
| 52422 | 10.43 | 3.53 | 5.39 | 1.44 | 2.32 | 0.67 | 0.80 | 9.07 | 0.85 | 1.07 | 1.35 | 1.12 | 3.86 |
| 52431 | 7.85 | 3.45 | 4.96 | 1.84 | 2.30 | 1.39 | 1.11 | 6.38 | 1.15 | 1.28 | 1.45 | 1.39 | 4.15 |
| 52440 | 7.45 | 2.33 | 10.18 | 1.63 | 1.79 | 1.00 | 0.77 | 8.84 | 1.19 | 1.86 | 2.53 | 0.73 | 2.71 |
| 52450 | 16.24 | 9.63 | 10.69 | 4.33 | 6.50 | 3.27 | 2.81 | 29.37 | 2.67 | 2.68 | 2.74 | 2.41 | 9.53 |
| 52460 | 4.53 | 1.92 | 4.40 | 2.17 | 2.02 | 1.54 | 1.63 | 2.20 | 1.69 | 1.78 | 1.85 | 1.94 | 2.14 |
| 52470 | 7.12 | 2.28 | 4.80 | 2.75 | 2.03 | 2.62 | 1.85 | 6.64 | 1.71 | 1.75 | 1.84 | 1.52 | 2.71 |
| 52485 | 11.64 | 3.87 | 6.72 | 2.51 | 3.40 | 1.90 | 1.48 | 5.91 | 1.14 | 1.17 | 1.37 | 1.87 | 6.31 |
| 52491 | 5.68 | 2.19 | 3.82 | 1.30 | 1.50 | 0.46 | 0.44 | 5.33 | 0.67 | 0.99 | 1.27 | 0.55 | 1.42 |
| 52500 | 24.90 | 6.09 | 16.97 | 8.42 | 11.71 | 1.53 | 1.69 | 12.12 | 1.93 | 2.19 | 2.45 | 2.35 | 13.52 |
| 52610 | 16.35 | 8.97 | 24.97 | 4.40 | 5.64 | 5.34 | 4.46 | 31.91 | 4.16 | 3.91 | 3.71 | 5.01 | 5.03 |
| 52620 | 4.71 | 4.42 | 5.25 | 3.49 | 4.34 | 1.06 | 1.03 | 4.81 | 1.05 | 1.04 | 1.04 | 0.78 | 4.81 |
| 52630 | 9.49 | 2.48 | 2.71 | 1.48 | 2.31 | 0.64 | 1.05 | 2.27 | 1.11 | 1.13 | 1.14 | 1.51 | 2.19 |
| 52700 | 6.94 | 2.46 | 9.84 | 3.11 | 4.84 | 2.53 | 1.76 | 4.89 | 1.44 | 1.42 | 1.54 | 2.20 | 2.23 |
| median (rrmse) | 8.67 | 3.28 | 5.11 | 2.55 | 2.44 | 1.53 | 1.55 | 5.63 | **1.31** | 1.54 | 1.64 | 1.56 | 3.55 |
| mean (rrmse) | 9.86 | 3.78 | 7.55 | 3.08 | 3.98 | 1.95 | 1.60 | 9.49 | **1.58** | 1.76 | 1.98 | 1.67 | 4.17 |

Table S8: Summary information on relative bias (from Table S6) and relative rmse (from Table S7) for the models listed in Table 1 using the full population but the simpler model (5). The best-performing models are indicated in bold.

|  | mean rb | mean abs(rb) | mean rrmse |
|---|---|---|---|
| Direct (HT) | **0.02** | 0.34 | 9.86 |
| Direct (GREG) | 0.49 | 0.93 | 3.78 |
| EBLUP | -1.75 | 4.30 | 7.55 |
| EBLUP (var = f(SC)) | 0.62 | 1.51 | 3.08 |
| pseudo-EBLUP | 0.97 | 1.65 | 3.98 |
| Robust synthetic | 0.26 | 1.76 | 1.95 |
| M-quantile naïve ($b_\psi = 1.345$) | **0.01** | 1.22 | 1.60 |
| M-quantile bias-adjusted ($b_\psi = 1.345$) | -0.41 | 5.68 | 9.49 |
| M-quantile robustly bias-adjusted ($b_\psi = 1.345$, $b_\phi = 1$) | -0.11 | 0.95 | **1.58** |
| M-quantile robustly bias-adjusted ($b_\psi = 1.345$, $b_\phi = 2$) | -0.20 | 0.98 | 1.76 |
| M-quantile robustly bias-adjusted ($b_\psi = 1.345$, $b_\phi = 3$) | -0.30 | 1.06 | 1.98 |
| weighted M-quantile naïve ($b_\psi = 1.345$) | -0.41 | 1.22 | 1.67 |
| weighted M-quantile bias-adjusted ($b_\psi = 1.345$) | **0.02** | **0.15** | 4.17 |

## S4 Investigation of the second tuning constant in the robustly bias-adjusted M-quantile estimator

Table S9: Relative rmse for robustly bias-adjusted M-quantile estimator with tuning parameter for the bias adjustment from 0.25 to 3 in steps of 0.25. The max – min summary row shows the range of the rmse over the considered values of $b_\phi$. The minimum rrmse over the range of $b_\phi$ values is highlighted in bold. (This is a version of Table 5 in the main paper, but showing results for all industry domains.)

| $b_\phi$ | 52110 | 52120 | 52200 | 52310 | 52321 | 52330 | 52413 | 52422 | 52431 | 52440 |
|---|---|---|---|---|---|---|---|---|---|---|
| 0.25 | 0.84 | **1.60** | 0.45 | **0.65** | 1.81 | **2.70** | 2.30 | 0.96 | 1.17 | **0.79** |
| 0.50 | 0.81 | 1.61 | 0.43 | 0.68 | 1.75 | 2.78 | **2.14** | 0.95 | 1.17 | 0.91 |
| 0.75 | **0.80** | 1.61 | **0.43** | 0.71 | **1.75** | 2.90 | 2.17 | 0.92 | 1.16 | 1.06 |
| 1.00 | 0.80 | 1.62 | 0.44 | 0.74 | 1.78 | 3.02 | 2.33 | **0.89** | **1.15** | 1.21 |
| 1.25 | 0.81 | 1.63 | 0.45 | 0.75 | 1.83 | 3.16 | 2.57 | 0.90 | 1.16 | 1.36 |
| 1.50 | 0.83 | 1.65 | 0.47 | 0.75 | 1.88 | 3.32 | 2.88 | 0.94 | 1.19 | 1.51 |
| 1.75 | 0.84 | 1.66 | 0.48 | 0.75 | 1.93 | 3.49 | 3.22 | 1.00 | 1.22 | 1.67 |
| 2.00 | 0.86 | 1.68 | 0.50 | 0.75 | 1.98 | 3.66 | 3.57 | 1.07 | 1.27 | 1.84 |
| 2.25 | 0.88 | 1.70 | 0.51 | 0.74 | 2.03 | 3.83 | 3.93 | 1.14 | 1.31 | 2.00 |
| 2.50 | 0.91 | 1.71 | 0.53 | 0.73 | 2.09 | 4.00 | 4.28 | 1.21 | 1.36 | 2.16 |
| 2.75 | 0.93 | 1.73 | 0.54 | 0.73 | 2.14 | 4.17 | 4.63 | 1.28 | 1.40 | 2.31 |
| 3.00 | 0.96 | 1.75 | 0.55 | 0.72 | 2.19 | 4.35 | 4.96 | 1.35 | 1.45 | 2.46 |
| max - min | 0.16 | 0.15 | 0.12 | 0.10 | 0.45 | 1.65 | 2.82 | 0.46 | 0.29 | 1.67 |

(cont)

| $b_\phi$ | 52450 | 52460 | 52470 | 52485 | 52491 | 52500 | 52610 | 52620 | 52630 | 52700 |
|---|---|---|---|---|---|---|---|---|---|---|
| 0.25 | 2.65 | **1.75** | 1.68 | 1.50 | **0.49** | **2.11** | 4.88 | **0.89** | 1.47 | 1.89 |
| 0.50 | 2.62 | 1.75 | 1.65 | 1.37 | 0.50 | 2.16 | 4.79 | 0.90 | 1.48 | 1.78 |
| 0.75 | **2.62** | 1.76 | **1.64** | 1.27 | 0.55 | 2.21 | 4.71 | 0.90 | 1.49 | 1.69 |
| 1.00 | 2.62 | 1.76 | 1.65 | 1.20 | 0.61 | 2.26 | 4.63 | 0.91 | 1.49 | 1.62 |
| 1.25 | 2.62 | 1.77 | 1.66 | 1.16 | 0.68 | 2.30 | 4.56 | 0.91 | 1.48 | 1.57 |
| 1.50 | 2.63 | 1.78 | 1.68 | **1.15** | 0.75 | 2.34 | 4.49 | 0.92 | 1.47 | 1.54 |
| 1.75 | 2.64 | 1.80 | 1.70 | 1.16 | 0.83 | 2.39 | 4.42 | 0.92 | 1.46 | **1.52** |
| 2.00 | 2.66 | 1.82 | 1.72 | 1.19 | 0.91 | 2.43 | 4.36 | 0.93 | 1.45 | 1.53 |
| 2.25 | 2.68 | 1.84 | 1.74 | 1.23 | 0.99 | 2.47 | 4.30 | 0.94 | 1.44 | 1.54 |
| 2.50 | 2.70 | 1.85 | 1.76 | 1.27 | 1.07 | 2.51 | 4.24 | 0.94 | 1.43 | 1.56 |
| 2.75 | 2.72 | 1.87 | 1.79 | 1.32 | 1.14 | 2.56 | 4.19 | 0.95 | 1.42 | 1.58 |
| 3.00 | 2.73 | 1.89 | 1.82 | 1.37 | 1.20 | 2.60 | **4.14** | 0.95 | **1.42** | 1.60 |
| max - min | 0.12 | 0.14 | 0.17 | 0.35 | 0.72 | 0.49 | 0.74 | 0.07 | 0.07 | 0.37 |

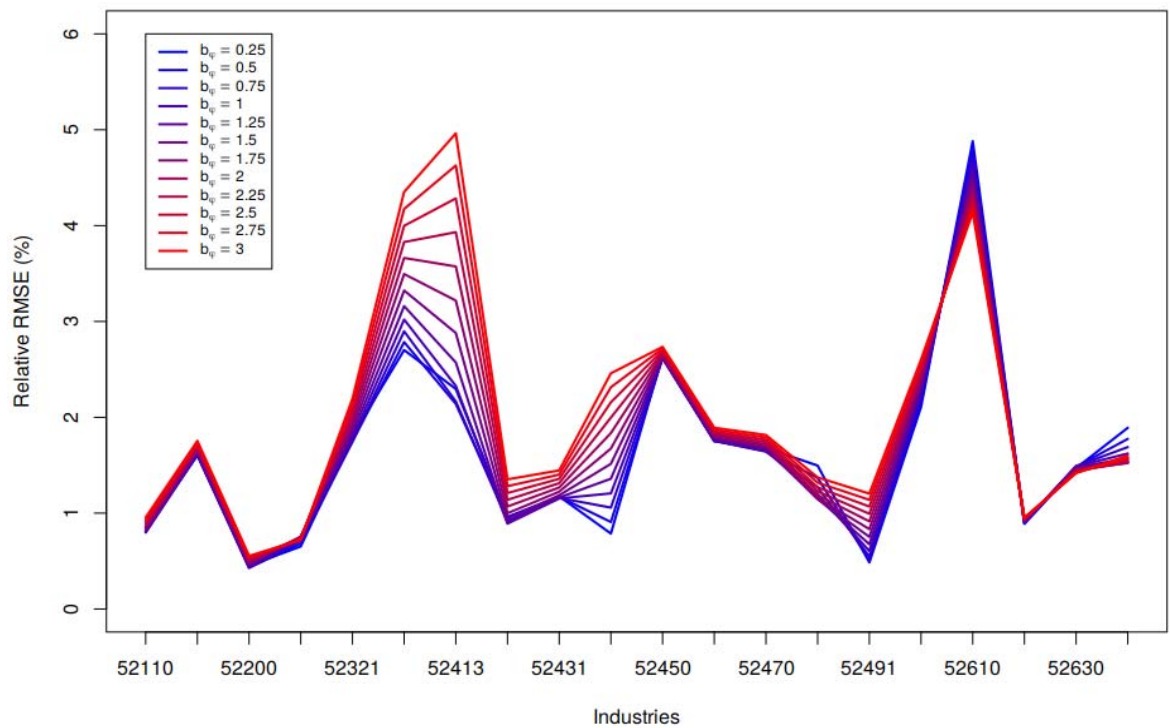

Fig. S2: The effect of changing $b_\phi$ on the relative rmse of the different industries. Note that some industries are much more affected than others, and the effects do not always go in the same direction.